\documentclass[useAMS,usenatbib]{mn2e}
\usepackage{graphicx}

\def\deg{\ifmmode^\circ\else$^\circ$\fi}

\def\msun{M$_{\odot}$ }

\def\micron{$\mu$m }
\def\hii{\mbox{H\,{\sc ii}}}
\def\neii{\mbox{[Ne\,{\sc ii}] }}

\title[Star formation in  G351.63-1.25]{Star formation activity in the southern Galactic 
H\,{\Large\bf II}~region G351.63-1.25}
\author[S. Vig et al.]{S. Vig,$^{1,2}$\thanks{E-mail: sarita@iist.ac.in} 
S. K. Ghosh,$^{2,3}$ D. K. Ojha,$^2$ 
R. P. Verma$^2$\thanks{Present address: 301, Vigyan, Plot 23, Sector 17, Vashi, Navi Mumbai 400703, India} 
and M. Tamura$^4$ \\
$^1$Department of Earth and Space Science, Indian Institute of Space Science and Technology, Thiruvananthapuram
 695 547, India \\
$^2$Tata Institute of Fundamental Research, Mumbai 400 005, India\\
$^3$National Centre for Radio Astrophysics, TIFR, Pune, 411 007, India\\
$^4$National Astronomical Observatory of Japan, Mitaka, Tokyo 181-8588, Japan}

\begin{document}

\date{}

\pagerange{\pageref{firstpage}--\pageref{lastpage}} \pubyear{}

\maketitle

\label{firstpage}

\begin{abstract}
The southern Galactic high mass star-forming region, G351.63-1.25, is a 
H\,{\sc ii}~region-molecular cloud complex
with a luminosity of $\sim 2.0\times10^5$~L$_\odot$, located at a distance of 2.4 kpc from the Sun.
In this paper, we focus on the investigation of the associated H\,{\sc ii}~region, embedded cluster and the 
interstellar medium in the vicinity of G351.63-1.25. We address the identification of exciting source(s) as well as the census of 
the stellar populations, in an attempt to unfold star formation activity in this region.
The ionised gas distribution has been mapped using the Giant Metrewave Radio Telescope (GMRT), India at three 
frequencies: 1280, 610 and 325 MHz. 
The \hii~region shows an elongated morphology and the 1280 MHz map comprises six resolved high density regions encompassed 
by diffuse emission spanning 1.4$\times$1.0~pc$^2$.
Based on measurements of flux densities at multiple radio frequencies, the brightest ultracompact core has electron temperature 
$T_e\sim7647\pm 153$~K and emission measure, $EM\sim2.0\pm0.8\times10^7$~cm$^{-6}$pc. The zero age main-sequence (ZAMS) 
spectral type of the brightest radio core is O7.5. We have carried out near-infrared observations in the JHK$_s$ 
bands using the SIRIUS instrument on the 1.4 m Infrared Survey Facility (IRSF) telescope. The near-infrared 
images reveal the presence of a cluster embedded in nebulous fan-shaped emission. The log-normal slope of the
K-band luminosity function of the embedded cluster is found to be $\sim0.27 \pm0.03$ and the fraction of the near-infrared
 excess stars is estimated to be 43\%. These indicate that the age of the cluster is consistent with $\sim 1$~Myr. Other 
available data of this region show that the warm (mid-infrared) and cold (millimetre) dust emission peak at different 
locations indicating progressive stages of star formation process. The champagne flow model from a flat, thin molecular 
cloud is used to explain the morphology of radio emission with respect to the millimetre cloud and infrared brightness.
\end{abstract}


\section{Introduction}
Our understanding of the formation of massive stars is poor relative to that of low-mass stars \citep{1987ARA&A..25...23S} although considerable 
progress is being made \citep{1999PASP..111.1049G, 2007prpl.conf..165B}. 
This is because the formation and early evolution of stars progress deep within the 
parental molecular cloud of gas and dust. Further adding to the difficulty, is the fast pre-main-sequence evolution 
of massive stars, their rarity as well as large distances (kpc scale or larger) as compared to their low mass counterparts. 
In addition, it has been observed that massive stars  
usually form in clusters or complexes, i.e. accompanied by swarms of stars 
of different masses \citep{1997A&A...320..159T}. The detailed study of massive star-forming complexes 
necessitates an investigation in different wavelength bands, in order to probe 
the distinct characteristics of the star formation process. Multiwavelength 
observations, therefore, hold the key to unraveling the least understood 
facets of high-mass star formation. 

Star forming complexes in the southern sky have been relatively less studied compared to the northern regions
and in this paper we investigate one such star forming region in detail.
The massive star-forming region, G351.63-1.25, (associated with
IRAS 17258-3637) is a \hii~region-molecular cloud complex with a luminosity of $1.9\times10^5$~L$_\odot$ 
\citep{2004A&A...426...97F}. We have adopted a distance of 2.4 kpc based on the studies by \citet{2000MNRAS.317..315V},
\citet{2004A&A...426...97F}, and \citet{2005A&A...440..121B}.
Millimetre continuum emission from cold dust in this region at 1.2 mm \citep{2004A&A...426...97F} shows the presence of a single dust core 
with total mass of $\sim1400$ M$_\odot$.
 \citet{2011ApJS..195....8C} have detected emission towards G351.63-1.25 at 2 and 3 mm and this is included in their 
QUaD Galactic Plane survey catalog. 
Molecular line investigation of this region has revealed a HC$_3$N core 
with tentative CO and SiO outflows \citep{2004AJ....128.2374S}. In the radio continuum, G351.63-1.25 has a very compact 
source \citep[FWHM$\sim6''$ at 3.7 cm;][]{1974ApJ...192..343B} which is surrounded by a more extended source 
\citep{1987A&A...171..261C}. More recent high resolution radio continuum observations at 8.7 GHz \citep{1998MNRAS.301..640W}
reveal the central region ($6''\sim0.07$~pc) to be irregular shaped with local peaks. 
\citet{1990ApJ...353..564G} have presented far-infrared observations of this region in
the 120-300 $\mu$m band using the Tata Institute of Fundamental Research (TIFR) 
100 cm balloon borne telescope. They have constructed a spectral energy distribution from 2 $\mu$m to 1 mm and
carried out simple radiative transfer calculations using a spherically
symmetric dust shell, for an assumed distance of 5 kpc. 
An infrared cluster located in this region has been discovered by \citet{2003A&A...404..223B} 
using the Two-Micron All Sky Survey (2MASS) survey. Further, high resolution 
K-band spectra of three young stellar objects located in this region have 
been obtained by \citet{2005A&A...440..121B, 2006A&A...455..561B} as a part of their survey to study 
massive young stellar objects. Methanol maser emission \citep{1994MNRAS.268..464S, 2000MNRAS.317..315V}, one of the 
signposts of massive star formation, has also been detected in this region. 

While few studies have looked at this massive star-forming region in general, there is no study focussing on the associated embedded cluster. Some 
of the questions that we aim to address relate to the identification of exciting source(s), census of the stellar populations,
and compilation of the available observations to construct a picture of the star formation activity in this region. 
To carry out this investigation, we have used a combination of infrared and radio 
wavebands. We probe the young cluster using deeper and high resolution near-infrared (NIR) observations as compared 
to the previous studies (eg. based on 2MASS). Our new low frequency radio observations of this \hii~region have the 
advantage that we can simultaneously image the compact as well as diffuse emission with moderate to high 
angular resolution.

The layout of the paper is as follows. In Sect. 2, we present the radio and NIR
observations as well as a description of other available data-sets used in this 
study. Sections 3-6 describe the results and in Sect. 7, we discuss the multiwavelength 
scenario for star formation in the light of various observational results. The 
conclusions are presented in Sect. 8.

\section{Observations and data reduction}

\subsection{Radio continuum observations using GMRT}

We have carried out low frequency radio continuum observations of G351.63-1.25 at 1280, 610 and 325 MHz using the 
Giant Metrewave Radio Telescope (GMRT) located at Khodad, India. The GMRT has a 
``Y" shaped hybrid configuration of 30 antennas, each of 45 m diameter. Six 
antennas are placed along each of the three arms (east, west and south) 
and twelve antennas are located in a random pattern within a compact area
$\sim 1\times1$ km$^2$ at the centre \citep{1991CuSc...60...95S}. The baselines 
($\sim$ 100 m - 25 km) provide sensitivity to large scale ($\sim5'$) 
diffuse emission as well as high angular resolution ($\sim4-20\arcsec$). The flux 
calibrators used during observations were 3C286 \& 3C48, and the phase 
calibrators used were 1626-298, 1822-096 and 1830-360. The details of the radio 
observations are listed in Table~\ref{radobs}. 

The National Radio Astronomy Observatory (NRAO) Astronomical Image Processing 
System (AIPS) was used for data reduction. The data were carefully checked for RF
interference and instrumental problems and suitably 
edited. The calibrated data are Fourier transformed and deconvolved using the IMAGR 
task in AIPS. Self calibration was carried out to remove the residual effects of
atmospheric and ionospheric phase corruptions and obtain the improved maps.
 In absence of automatic gain control of the antenna system, it becomes important to correct for the 
system temperature for target sources in 
the Galactic plane, particularly at lower frequency bands (610 and 325 MHz) where the contribution of 
sky temperature to system temperature is significant. 
The fluxes at these frequencies therefore need to be scaled by a factor obtained in the following way.
A system temperature for each frequency, $T_{corr}$, was obtained using the sky temperature towards the target source from the map of 
\citet{1982A&AS...47....1H} at 408 MHz. A correction factor, $(T_{corr} + T_{sys})/T_{sys}$, has been used to scale the deconvolved images, 
where $T_{sys}$ is the system temperature for sources (flux calibrator in our case) away from the Galactic plane. 
The radio brightness distribution from ionised gas shows a number of high density regions which have been 
characterised using the task JMFITS (in AIPS). This task fits a 2D-Gaussian function to a selected brightness 
distribution (having S/N $>10$ at 1280 MHz) and gives the position of peaks and the corresponding 
flux densities.

\subsection{Near-infrared observations using IRSF}

The region associated with G351.63-1.25 was imaged in the NIR
broad bands $J$ (1.25  $\mu$m), $H$ (1.63 $\mu$m), and
$K_{\rm s}$ (2.14  $\mu$m) on 11 July 2004 using the instrument SIRIUS on the 
1.4 m InfraRed Survey Facility (IRSF) telescope. SIRIUS is a three-color simultaneous camera equipped with 
three 1024 $\times$ 1024 HgCdTe arrays. The imaging observations were 
centered on $\alpha_{2000} = 17^{h}29^{m}17^{s}$,
$\delta_{2000} = -36^{\circ}40^{\prime}13^{\prime\prime}$.
The field-of-view in each band is $\sim$ $7.8\arcmin$
$\times$ $7.8\arcmin$, with a pixel scale of $0.45\arcsec$ at
the f/10 Cassegrain focus. Further details of the instrument are given in
\citet{1999sf99.proc..397N} and \citet{2003SPIE.4841..459N}. 
We obtained 90 dithered frames each of 10s, giving a total
integration time of 900s in each band. The observing conditions were
photometric and the average FWHM during the observing period was $\sim$
$1.1\arcsec-1.4\arcsec$. Data reduction was carried out using a
software pipeline based on IRAF package tasks. Dome flat-fielding and sky
subtraction with a median sky frame were applied. Photometry of point
sources was performed using the point spread function (PSF) algorithm ALLSTAR in the DAOPHOT
package \citep{1987PASP...99..191S} within the IRAF environment. The PSF was determined
from 20 to 27 relatively bright and isolated stars of the field. For every band (J, H 
and K$_s$), we have used
an aperture radius of 1 FWHM,  with appropriate aperture corrections for the final photometry.

For photometric calibration, we used 21 isolated sources from the 
2MASS Point Source Catalog (PSC). 
 The 2MASS covers the sky in the three NIR broad bands $J$, $H$ and K$_s$.
The sources used for photometric calibration have $K_s$ magnitudes 
lying in the magnitude range 11-16 and having $K_s$ magnitude errors $<$ 0.1 in all the three, J, H and K$_s$ bands. 
These 2MASS sources were also used for absolute position
calibration and a position accuracy better than $\pm0.1\arcsec$
was achieved. A comparison with 2MASS sources indicates that the brighter IRSF
sources with $K_s<11$ mag are saturated. The magnitudes of these sources 
were replaced with the corresponding magnitudes from the 2MASS PSC. 
For comparison with other studies, the magnitudes were transformed to the Bessel and Brett or BB system, \citep{1988PASP..100.1134B} 
 system using the relations given at the 2MASS
website\footnote{www.ipac.caltech.edu/2mass/releases/allsky/doc}.

\subsection{Other archival datasets}

In order to complement our NIR and radio study of this region, we have used the available infrared images and catalog 
data from the Wide-field Infrared Survey Explorer (WISE), Midcourse Space Experiment (MSX) and Infrared Astronomical Satellite
(IRAS) archives, the details of which are given below. The {\it Spitzer Space Telescope} archive shows that this 
region has been partially covered using the InfraRed Array Camera (IRAC) in two bands: 4.5 and 8.0 $\mu$m. Although of 
higher angular resolution, these images are saturated near the peak.

\subsubsection{\it WISE}

The Wide-field Infrared Survey Explorer \citep[WISE,][]{2010AJ....140.1868W} mapped the sky at NIR and mid-infrared (MIR) 
wavebands of 3.4, 4.6, 12, and 22 μm ($W1$, $W2$, $W3$ and $W4$ bands, respectively, hereafter) using a 40-cm cryogenically 
cooled telescope. The angular resolution achieved is  $\sim 6\arcsec$ at  3.4, 4.6 and 12 $\mu$m bands and 
$\sim12\arcsec$ at 22 $\mu$m band \citep{2010AJ....140.1868W}. The region associated with G351.63-1.25 is saturated at the 
longer MIR wavebands, in $W3$ and $W4$ bands. Hence, for the present study, we have only considered $W1$ and $W2$ bands. 
The images and catalog sources were taken from the All Sky Data Release\footnote{http://wise2.ipac.caltech.edu/docs/release/allsky/}. 

The catalog sources in the vicinity of G351.63-1.25 have extended source flag of 2 or 3. This indicates that the WISE
 profile-fit and standard aperture measurements, which are optimized for point sources (with extension flag = 0), 
systematically underestimate the true flux of these objects and larger aperture photometry is recommended. 
We have, therefore, carried out aperture photometry anew. A field of size, $10\arcmin\times10\arcmin$ 
 centred on $\alpha_{2000} = 17^{h}29^{m}16.1^{s}$, $\delta_{2000} = -36^{\circ}40^{\prime}7^{\prime\prime}$, 
in the WISE bands was considered. The PSF in these bands (W1 and W2) has been estimated using few bright isolated
 point sources and is found to be $9\arcsec$ and $9.6\arcsec$ at W1 and W2 bands, respectively. The DAOPHOT 
library of IDL has been used to carry out aperture photometry. Considering that the field is crowded near the
 region of interest, we have used aperture and sky annuli of $8\arcsec$, $11\arcsec$ and $27\arcsec$,
 respectively for both W1 and W2. The photometry was calibrated with few bright, isolated sources from the WISE catalog. 

The sources from the NIR IRSF catalog were compared with those in the WISE catalog by searching for counterparts within 
$2\arcsec$ search radius. This search radius has been selected considering the typical seeing conditions in near-infrared
observations.
The number of sources having IRSF counterparts in W1 and W2 bands are 170 and 99, respectively. 

\subsubsection{MSX}

The Midcourse Space Experiment (MSX) surveyed the entire Galactic plane within $|$b$|\le 5^{\circ}$ in four MIR wavebands: 8.3, 12.1, 14.7
and 21.3 \micron at a spatial resolution of $\sim18.3\arcsec$ \citep{2001AJ....121.2819P}. The 
panoramic images of the Galactic plane survey of MSX were taken
from IPAC ({\tt http://irsa.ipac.caltech.edu/applications/MSX/}). 
Although of lower angular resolution compared to WISE, the MSX images have been considered because the
required coverage of this region at longer wavelengths ($\lambda\ge 8$~$\mu$m) is unsaturated and complete.
The MSX maps were used to compare the distribution of dust with the ionised gas in this region. 
Further, point sources in the region associated with G351.63-1.25 have been selected from MSX Point 
Source Catalog Version 2.3 \citep{2003AAS...203.5708E}.

\subsubsection{IRAS-LRS}

G351.63-1.25 (as IRAS 17258-3637) appears in the Infrared Astronomical Satellite (IRAS) - Low Resolution Spectrometer (LRS) 
Catalog \citep{1986A&AS...65..607O}. Our aim in examining the IRAS-LRS spectrum ($8-22$ \micron) is to ascertain the 
presence and probe the \neii emission from the ionised gas in this \hii~region. 

\section{Ionised gas distribution}

\begin{figure*}
\includegraphics[scale=.80,angle=0]{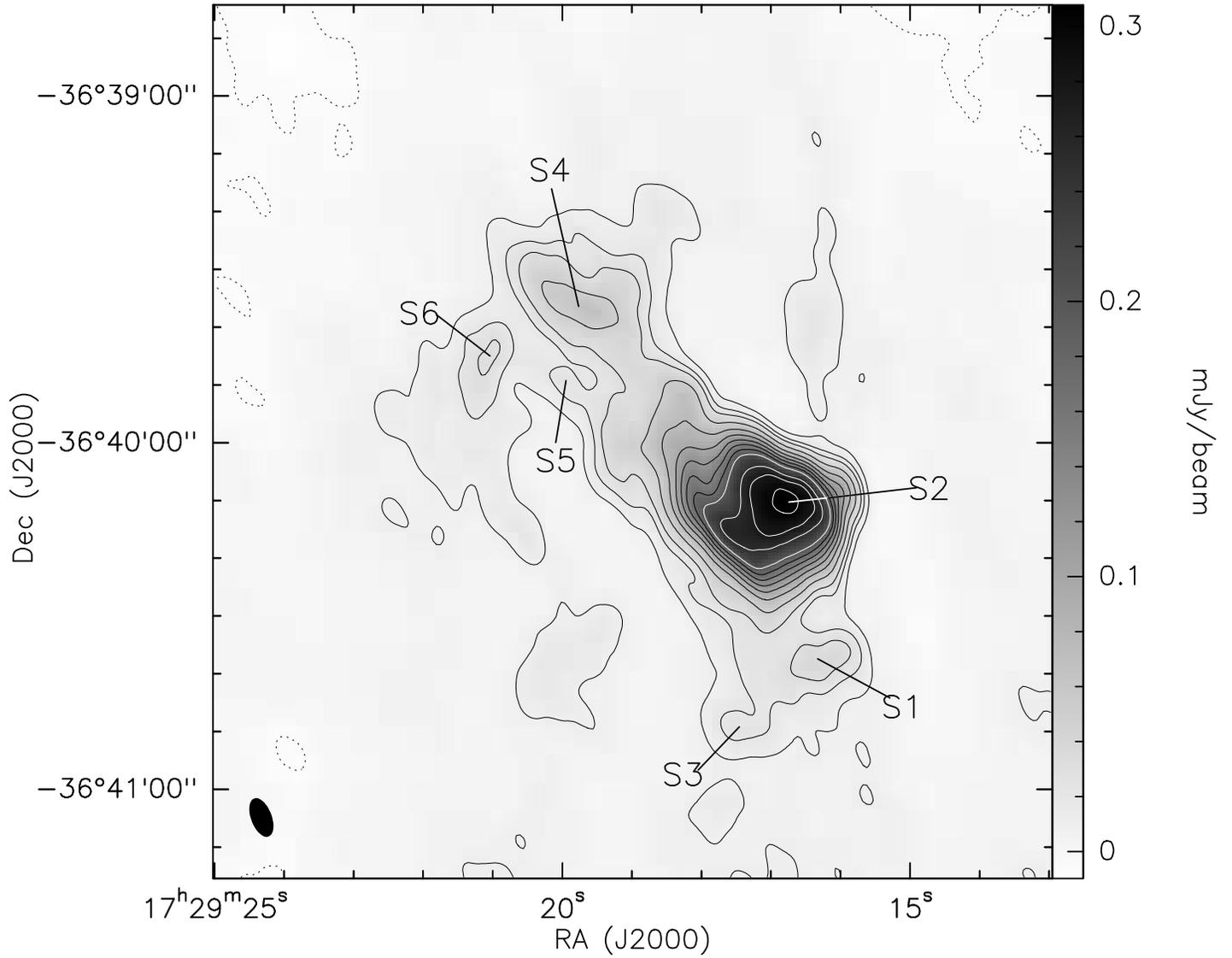}
\caption[GMRT maps of G351.63-1.25 at 1280 MHz]{Radio continuum emission at 
1280 MHz in the region around G351.63-1.25. The contour levels are at
 -5, 10, 20, 30, 50, 75, 100, 125, 150, 175, 200, 250, 275 and 300 mJy/beam. The negative contour is 
indicated by a dotted line.
The discrete radio sources are represented by the numbers as listed in 
Table~\ref{radsource}. The beam size is $7.0\arcsec \times3.6\arcsec$ (shown as dark ellipse in the lower left corner of the figure) and the rms noise 
is 2 mJy/beam.}
\label{g351_rad1}
\end{figure*}

\begin{figure*}
\includegraphics[scale=.35,angle=0]{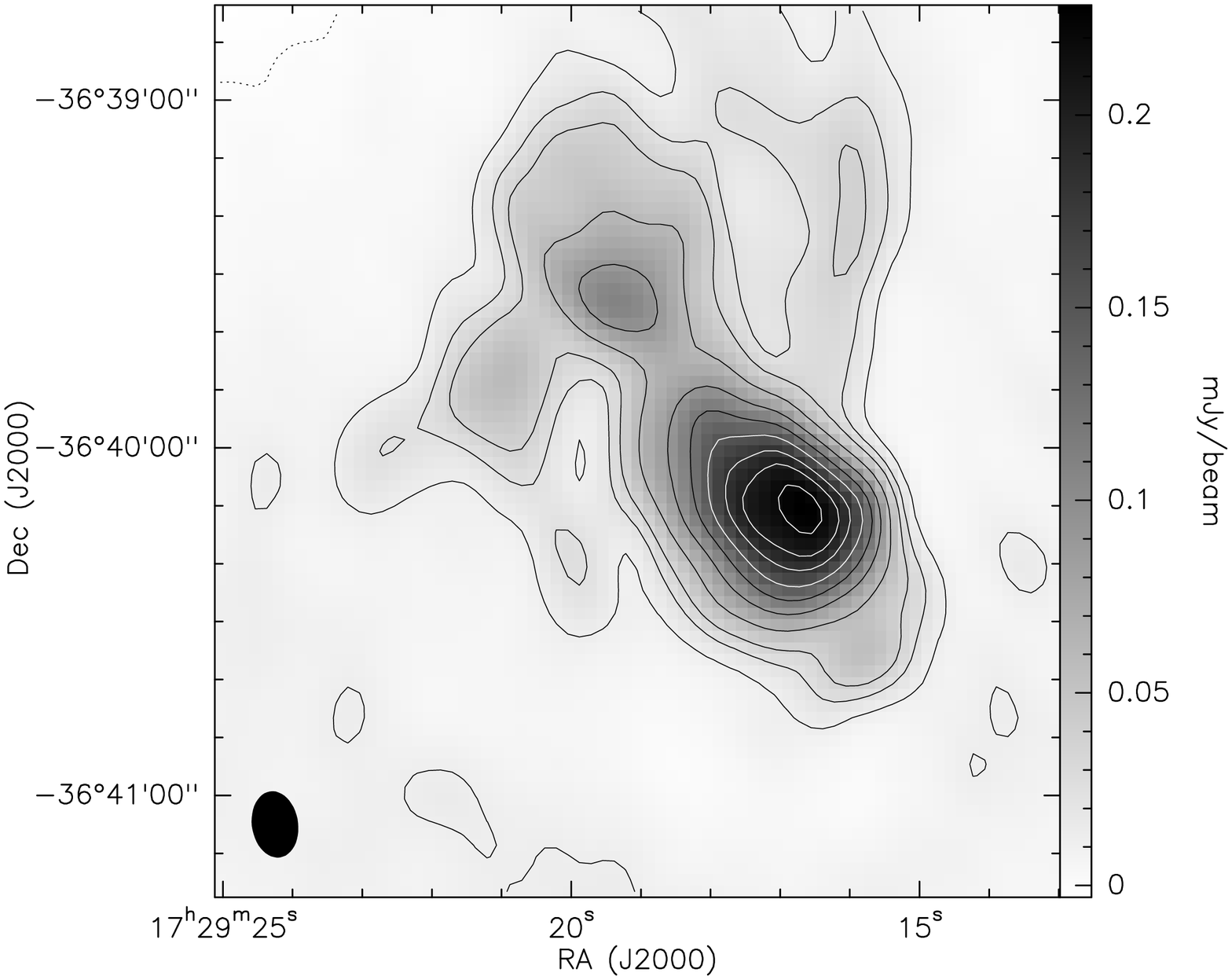}
\includegraphics[scale=.35,angle=0]{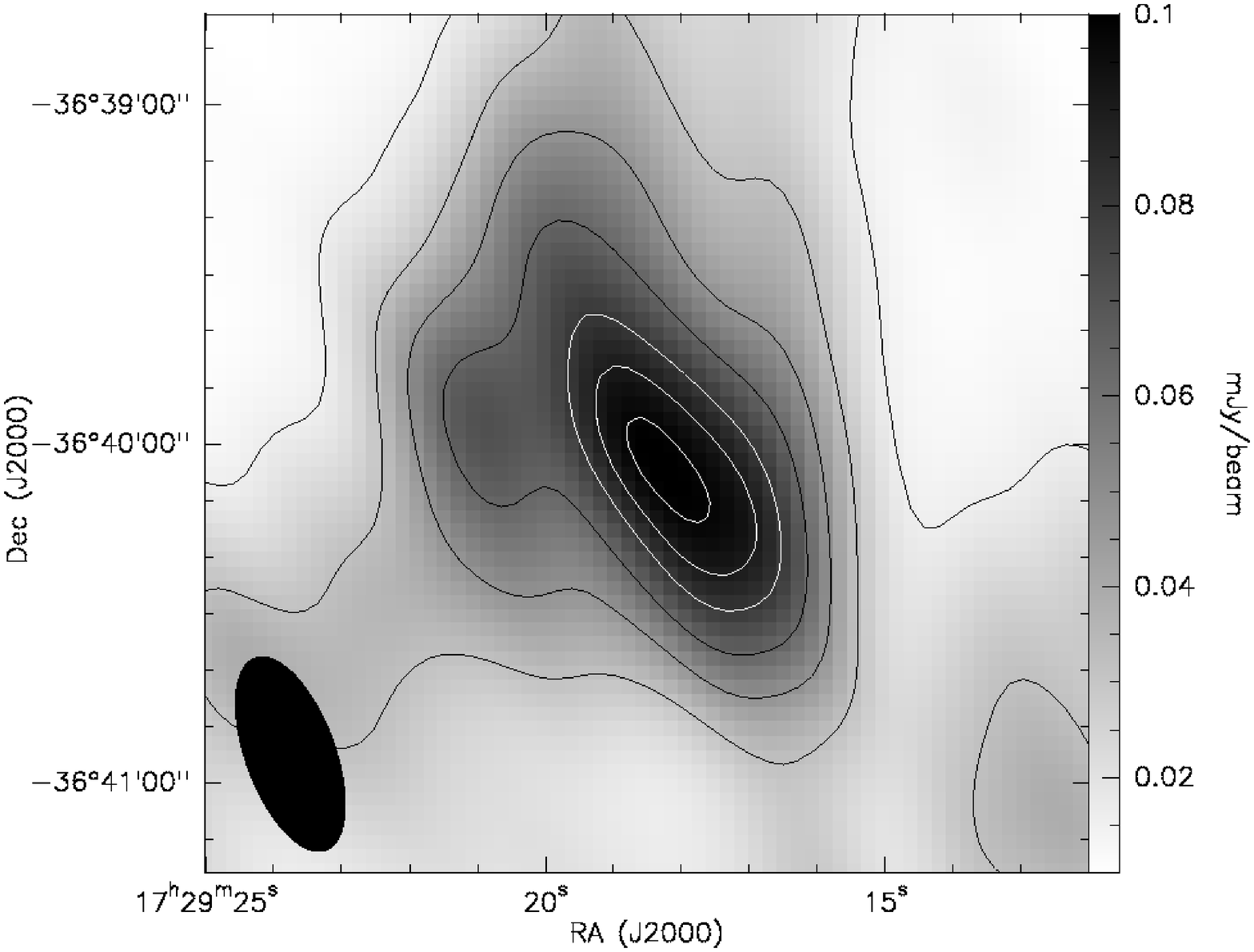}
\caption[GMRT maps of G351.63-1.25 at 610 and 325 MHz bands]
{Radio continuum emission at 610 (left) and 325 MHz
(right) in the region around G351.63-1.25. The contour levels are at
 -1, 12, 24, 36, 60, 90, 120, 150, 180, 200 and 220 mJy/beam (left) and 
16, 32, 48, 64, 80, 90 and 98 mJy/beam (right). The negative contour is represented by a dotted line. 
The corresponding beam is shown in the lower left corner of each image.}
\label{g351_rad2}
\end{figure*}

The radio continuum emission maps of the region around G351.63-1.25 
obtained for frequencies: 1280, 610 and 325 MHz are shown in 
Figs.~\ref{g351_rad1} and \ref{g351_rad2}. The synthesized beams and rms noise in 
these uniformly weighted images are listed in Table~\ref{radobs}. The highest resolution map at 
1280 MHz has a beam size $7.0\arcsec\times3.6\arcsec$. G351.63-1.25 is well resolved at all the three frequencies 
and the radio maps show diffuse emission in addition to several high density regions. The total 
flux densities in the maps up to $3\sigma$ level at
1280, 610 and 325 MHz are $10.83\pm0.06$ Jy, $6.11\pm0.03$ Jy and $3.86\pm0.02$
Jy, respectively. The size of the radio emitting gas, based on our interferometric observations 
($2.0\arcmin\times1.5\arcmin\sim1.4$~pc$\times1.0$~pc), samples most of the extended emission. This is evident when we compare the
  size of radio emitting gas from our maps with single dish measurements 
of this region by \citet{1987A&A...171..261C} at 5 GHz. These authors used the 64-m Parkes telescope and they detect 
emission up to $2.5\arcmin$. 

Six high density ionised regions (compared to the immediate surroundings) have been identified from the map at 
1280 MHz. These are labelled as S1, S2, ..., S6 in Fig.~\ref{g351_rad1}. At 610 MHz, we see four high density
 ionised regions (located close to S1, S2, S4 and S6) as compared to six, detected at 1280 MHz. 
 This is likely to be due to a combination 
of optical thickness effects as well as beam size effects (flux from the point source relative to the 
diffuse flux in the vicinity) in the images. Details of these regions along with their peak
 and integrated flux densities at 1280 and 610 MHz are
 listed in Table \ref{radsource}.  In the 325 MHz
band, we do not see any point-like source due to poor angular resolution in this band. 

\subsection{Physical conditions in S2}

\begin{figure*}
\vspace*{-0.5cm}
\includegraphics[scale=.50,angle=0]{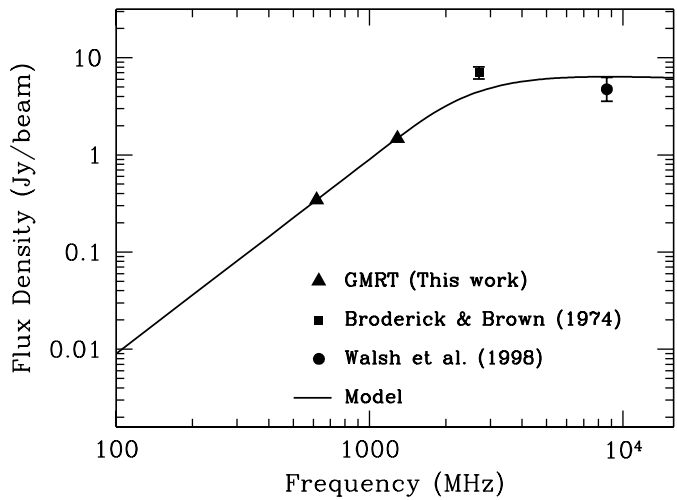}
\vspace*{-0.5cm}
\caption[]{Free-free emission from the core S2 in G351.63-1.25 at low frequency radio wavebands. The observed 
measurements (with errorbars) are represented by symbols while the solid line shows the modeled flux densities. 
The filled triangles are measurements from GMRT while the filled circle and square represent flux density measurements 
from \citet{1998MNRAS.301..640W} and \citet{1974ApJ...192..343B}, respectively. For GMRT measurements, the errors in 
flux are smaller than the symbol size used.
These flux densities correspond to a beam size of
$12\arcsec$.
}
\label{TEM}
\end{figure*}

We have estimated the electron temperature and emission measure of S2 using fluxes at 1280 and 610 MHz by modelling the radio emission to be 
free-free emission from isothermal, spherically symmetric and homogeneous ionised gas distribution \citep{1967ApJ...147..471M}. 
 Under these approximations, the flux density is given by  

$$S_\nu = 3.07\times10^{-2}T_e\nu^2\Omega(1-e^{-\tau(\nu)})$$
$$\tau(\nu)=1.643\times10^5\, a(\nu,T_e)\, \nu^{-2.1}\:(EM)\:T_e^{-1.35}$$

where $S_\nu$ is the integrated flux density in Jy, $T_e$ is electron temperature
in K, $\nu$ is the frequency in MHz, $\tau$ is the optical depth, $\Omega$ is
the solid angle subtended by the source in Steradian (which corresponds to the
synthesized beam size for the unresolved core, i.e $\Omega = 1.33 \; \theta_x \times \theta_y$ for a gaussian beam), and $EM$ is the emission
measure in cm$^{-6}$pc. The factor $a(\nu,T_e)$ corrects for approximation to the opacity. We take $a(\nu,T_e)=0.99$ obtained from Table 6 of
\citet{1967ApJ...147..471M} for the frequency range $0.6-8$ GHz for
$T_e=10,000$ K. In addition to fluxes at 1280 and 610 MHz, we have also 
used fluxes at 2.7 GHz \citep{1974ApJ...192..343B} and 8.64 GHz \citep{1998MNRAS.301..640W}. They  
find the presence of a compact \hii~region (of size $12\arcsec$ at 2.7 GHz and $6\arcsec$ 
at 8.64 GHz, respectively). A non-linear least-squares Marquardt-Levenberg
algorithm is used to fit the above equation to the flux densities of S2 at four frequencies 
(0.61, 1.28, 2.7 and 8.64 GHz), after convolving 
these flux densities to the lowest resolution among the four bands ($12\arcsec\times 12\arcsec$). 
The best-fit model flux densities along with the observed data points are shown in Fig.~\ref{TEM}. 
For this model, the electron temperature and emission measure (with 1$\sigma$ errors) are $7647\pm153$
 K and $2.0\pm0.8\times10^7$ cm$^{-6}$pc, respectively. This translates to an electron density $n_e = 1.2\pm0.8\times10^4$~cm$^{-3}$ for a 
size of 0.14 pc ($\sim12\arcsec$) assuming constant density. If we consider the flux density at 8.6 GHz by \citet{1974ApJ...192..343B} instead, we get a
similar $T_e$ value while the electron density is 41\% higher. In this paper, we proceed with our former value of $n_e$.
The electron temperature is consistent with $T_e = 7000$~K obtained by \citet{1987A&A...171..261C} using 
recombination line measurements. Considering that \hii~region is 7.4 kpc from the Galactic centre, this value of $T_e$ is in accordance 
with the variation of electron temperature with Galactocentric distance in ultracompact \hii~regions, due to a gradual change in the metallicity 
\citep{2002ARA&A..40...27C}. 

The electron density has been obtained by convolving beams to the lowest resolution beam at 610 MHz. Our 
highest resolution image at 1280 MHz has a beam size corresponding to $\sim 0.08$~pc$\times0.04$~pc. \citet{1974ApJ...192..343B}
 and \citet{1998MNRAS.301..640W} have obtained the bright (small angular scale) component of size 0.07 pc 
($\sim 6\arcsec$). This  implies that $n_e$ is likely to be $>1.2\times10^4$~cm$^{-3}$  in the bright region
 if we relax the uniform density assumption. The small size and high density imply that 
it is an ultracompact core embedded within an extended \hii~component \citep{2002RMxAC..12...16K}. In our case, 
we observe  several other high density ionised structures within a more diffuse component. Whether these structures 
represent a cluster of compact \hii~regions (each having an embedded exciting source) or high density 
clumps within the molecular cloud ionised by a single star is discussed in the later sections.

For free-free optically thin emission, the radio flux density is directly proportional to the flux of ionising photons. 
At 1280 MHz, S2 is optically thick, $\tau\sim 1.5$. Hence, we have used
the optically thin flux density at 8.64 GHz \citep{1998MNRAS.301..640W} to estimate the 
spectral type of the exciting source. We determine the excitation parameter, $U$
 \citep{1969ApJ...156..269S,1973AJ.....78..929P,1999ApJ...512..260J} 
which is given by the expression:

$$U = R_s\,n_e^{2/3} = 3.6304\,\left[ a(\nu, T_e)^{-1} \,\nu^{0.1} T_e^{0.35}\, S_\nu\,d^2 \right]^{1/3}$$  

Here, $U$ is in pc~cm$^{-2}$, $R_s$ is the Str\"{o}mgren radius, $d$ is the distance in kpc. The other terms are same as earlier. The flux of 
Lyman continuum photons is proportional to $U^3$. We obtain $U = 42.3$ which corresponds to a ZAMS star of spectral type O7.5 from the 
tables of \citet{1973AJ.....78..929P}. The error in flux density leads to a 
change of $\sim10$\% in $U$ that corresponds to a change of $< 0.5$ in spectral type. 
For $n_e = 1.2\times10^4$~cm$^{-3}$, the Str\"{o}mgren radius is $R_s = 0.08$~pc.

\section{Emission from dust}

\begin{figure*}
\hskip -1.cm
\includegraphics[scale=.4,angle=0]{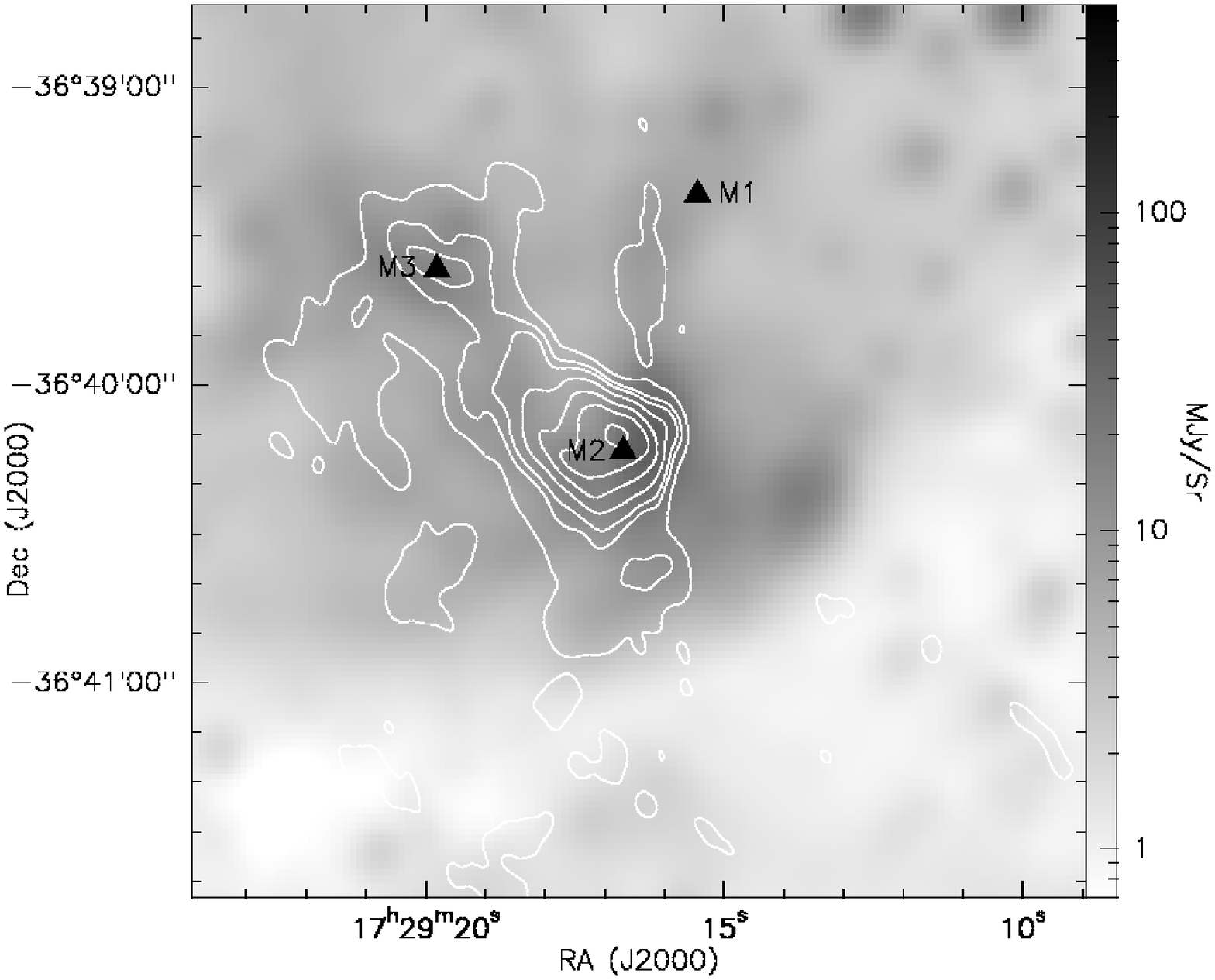}
\includegraphics[scale=.4,angle=0]{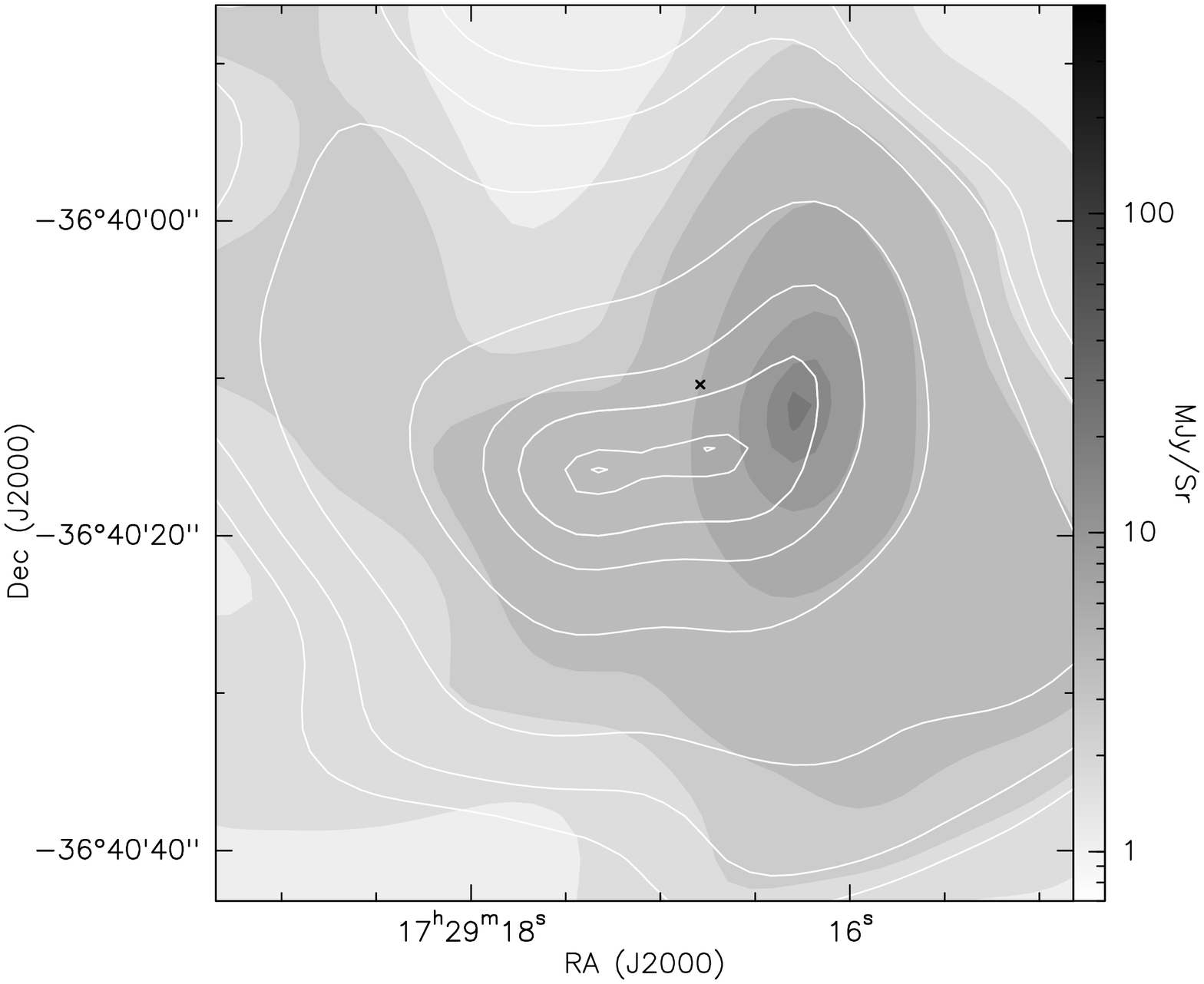}
\caption[WISE]{ The grayscale represents 3.4 $\mu$m WISE image towards G351.63-1.25. (Left) The contours represent 
radio continuum emission from ionised gas at 1280 MHz with levels at 10, 30, 50, 100, 180, 250 and 300 mJy/beam 
where beam size is $\sim7.0\arcsec\times3.6\arcsec$. The triangles represent MSX PSC sources. (Right) The contours represent emission 
from WISE for the central region at 4.6 $\mu$m with levels at 9.5, 11.3, 15, 30, 45, 52.5, 52.5, 58.7 and
 59.7 MJy/Sr. 
The cross represents the position of S2.}
\label{wise}
\end{figure*}

The emission from warm dust at 3.4 $\mu$m WISE 
 image in the vicinity of G351.63-1.25 is shown in Fig.~\ref{wise}~(left). 
In the image, emission in the form of diffuse filamentary structures is clearly perceived.  Although the distributions 
of diffuse emission at 3.4 and 4.6~$\mu$m  are similar, there is a difference in emission near the peaks.  While at 
3.4~$\mu$m we observe a single emission peak, two emission peaks are discerned at 4.6~$\mu$m shifted to the east. 
This can be seen in Fig.~\ref{wise}~(right). Note that the emission peaks at 3.4 and 4.6~$\mu$m are extended and hence
 point source fluxes  would be underestimated using photometry.
The MSX survey covers the longer wavelength ($\lambda\ge8\;\mu$m) MIR region 
and among the four bands, the best sensitivity is obtained in the 8.3 $\mu$m band.
The diffuse emission in MIR from warm dust at 8.3 $\mu$m follows a similar trend as the WISE NIR images, albeit with 
poorer resolution ($\sim18\arcsec$). A strong 
emission peak is observed at $\alpha_{2000}$ = $17^h$ $29^m$ $16.6^s$, 
$\delta_{2000}$ = -36$^{\circ}$ 40$\arcmin$ 15$\arcsec$. We refer to this 
as the MSX MIR peak. Another secondary MSX emission peak 
is detected towards the north-east at $\sim \alpha_{2000}$ = 
$17^h$ $29^m$ $20^s$, $\delta_{2000}$ = -36$^{\circ}$ 39$\arcmin$ 35$\arcsec$. 
However, the dominant contribution to this peak is from diffuse emission 
as can be seen  by comparison with the WISE images. 
Integrating the spectral energy distribution of G351.63-1.25, constructed using flux densities (from a 
circular region of $3\arcmin$
around the IRAS peak) from WISE, MSX, IRAS-HIRES maps at 12, 25, 60 and 100 $\mu$m \citep{1986A&AS...65..607O},
 the IRAS-LRS spectrum, 
TIFR map at 150 $\mu$m \citep{1990ApJ...353..564G} and  SIMBA map at 1.2 mm \citep{2004A&A...426...97F}, the total 
luminosity comes out to be $2\times10^5$ L$_\odot$.

From the MSX PSC, we find 3 sources within a circular region of 1$\arcmin$
centered on the radio peak S2. These are listed in Table~\ref{msxpsc}. We
have considered only those fluxes where the MSX quality flag is $\ge2$. The
quality flag is based on detection in various bands and the signal-to-noise
ratio. These sources are marked on the 3.4 $\mu$m 
WISE image in Fig.~\ref{wise}~(left) and labelled as M1, M2 and 
M3. M2 has the largest flux density and corresponds to the MSX 
MIR peak (mentioned earlier) while M3 corresponds to the secondary MSX peak. 
M1 and M3 appear to trace the filamentary diffuse structures in the WISE 
images. This has also been confirmed with the high resolution {\it Spitzer}-IRAC 8.0 $\mu$m image which partially covers 
this region.

\section{Embedded cluster}

\begin{figure*}
\includegraphics[scale=0.7,angle=0]{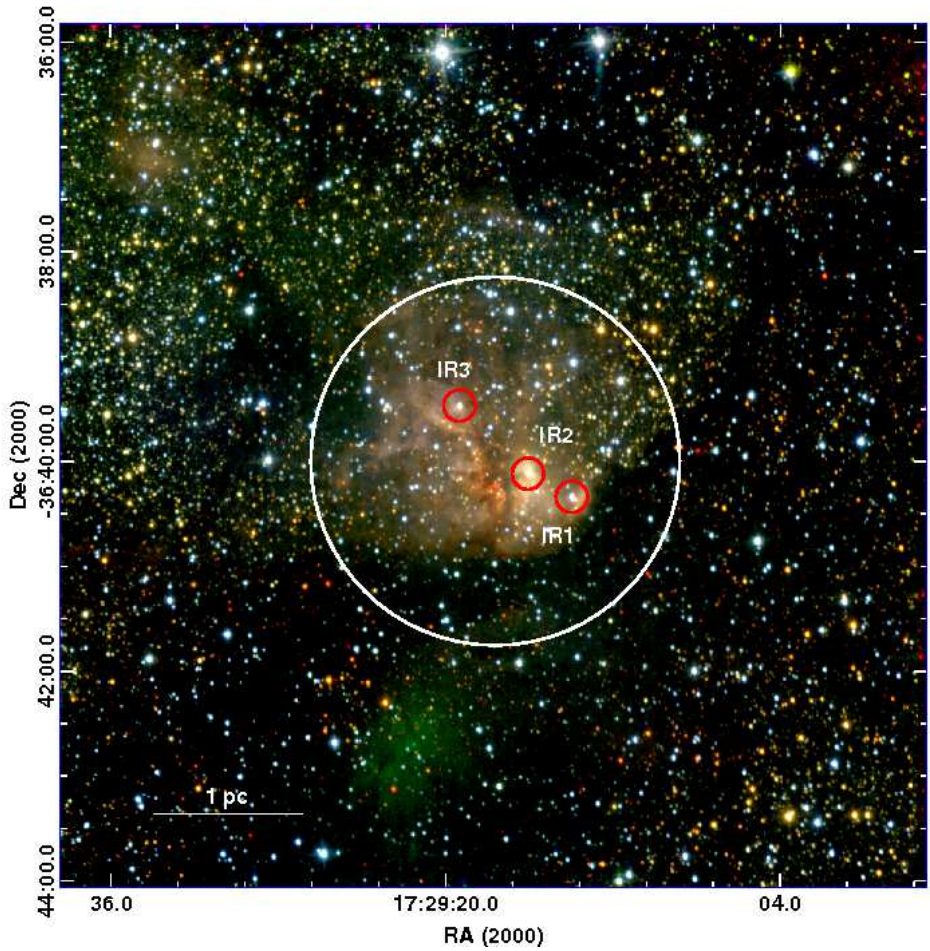}
\caption[JHK composite from IRSF]{JHK$_s$ colour-composite image of G351.63-1.25
where blue, green and red represent the emission in the J, H and K$_s$ bands, respectively. The 
size of the image is $7.8\arcmin\times7.8\arcmin$. North is 
up and east is towards the left. The circular region used for examining the 
embedded cluster is enclosed within the large white circle. Also shown are the three bright stellar objects 
 (IR1, IR2, IR3) whose spectral types are known. (A color version of this figure is 
available in the online journal.)}
\label{jhk}
\end{figure*}

The NIR images from IRSF have been used to investigate the embedded stellar cluster in this region. The $JHK_s$ 
colour-composite image 
of G351.63-1.25 region is shown in Fig.~\ref{jhk}. The colour-composite image 
shows fan-shaped nebular emission apart from a number of point sources. There is a sharp drop in diffuse intensity 
towards the lower edge of the fan-shaped emission and the reduction in number 
of point sources is prominent. Further south, faint diffuse emission can 
be discerned pointing towards the presence of a  high extinction ridge.
We also notice diffuse emission towards the north-east (top left corner).

\subsection{Nature of Stellar sources}

\begin{figure*}
\hskip -1.cm
\includegraphics[scale=.45,angle=0]{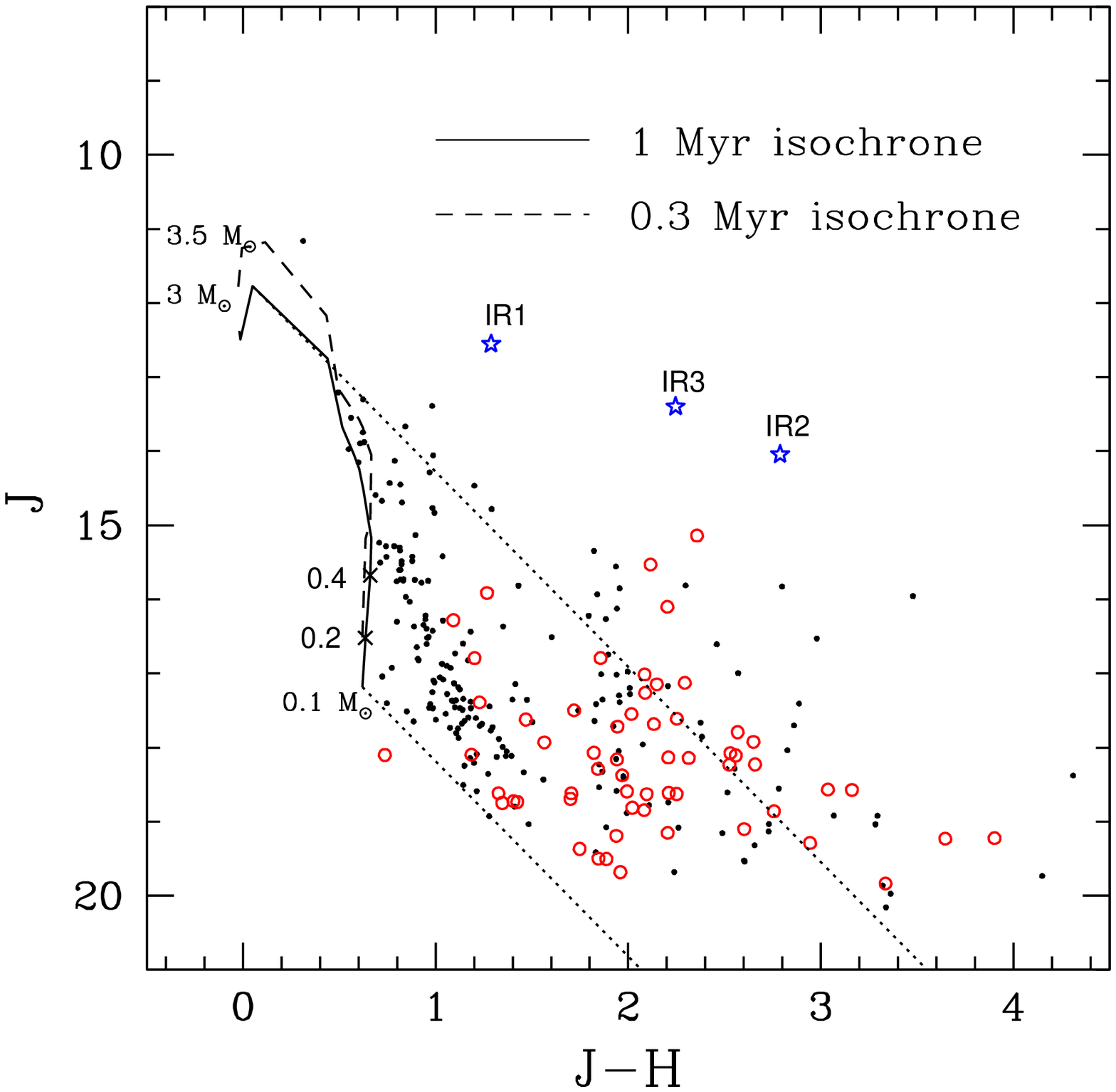}
\hskip -1.cm
\includegraphics[scale=.45,angle=0]{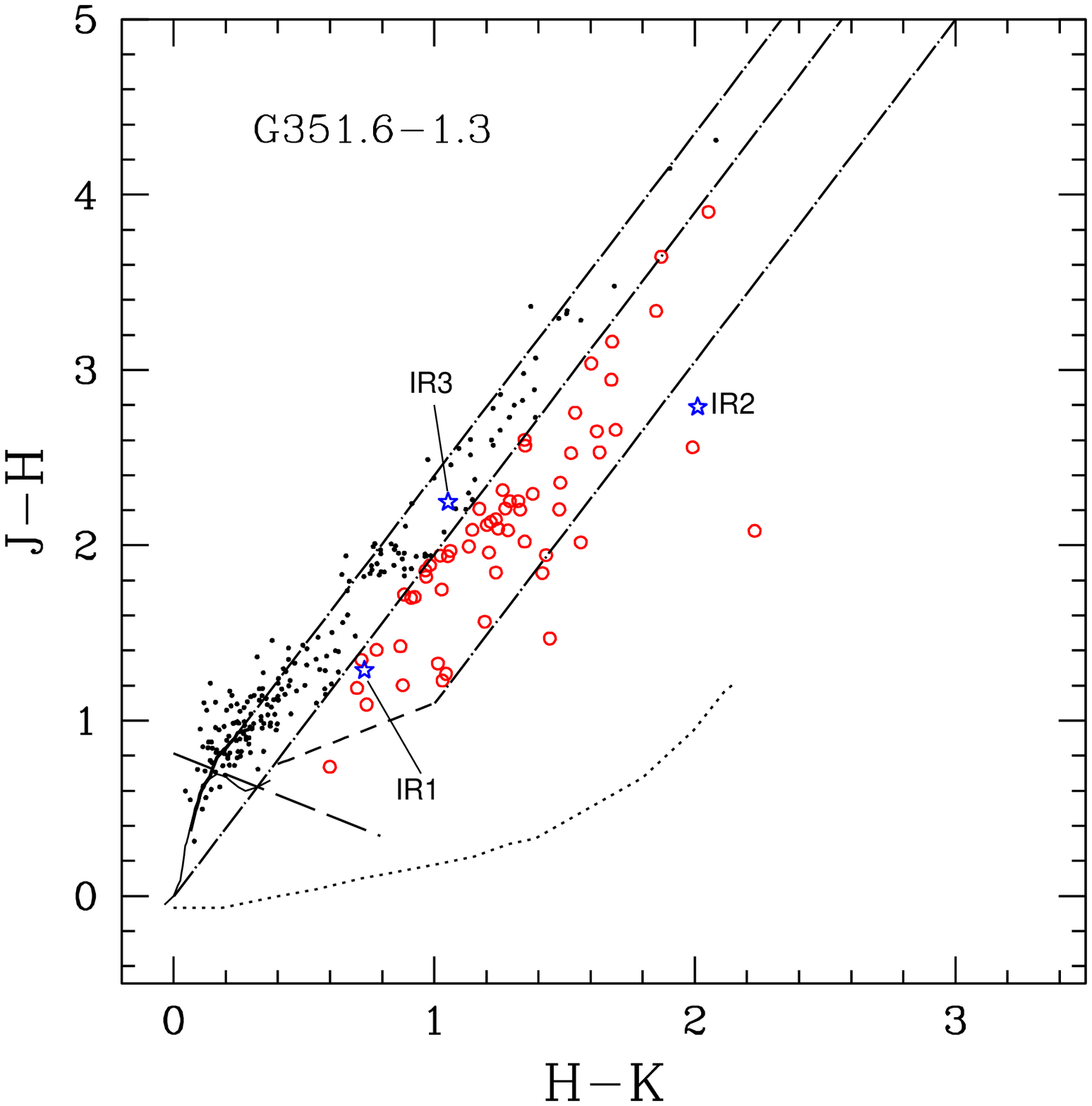}
\caption[CMD and CCD]{Colour-magnitude (left) and colour-colour (right)
diagrams for the sources detected in all the three JHK$_s$ bands using IRSF-SIRIUS 
in the region G351.63-1.25. Magnitudes of sources brighter than 11 mag in the K$_s$ band 
were replaced with the corresponding magnitudes from the 2MASS PSC. The open circles represent 
the sources showing an infrared excess in the colour-colour diagram. The asterisks represent 
young stellar objects whose spectral type has been inferred based on other spectroscopic studies. 
The rest of the sources are represented by dots.
In the colour-magnitude diagram, the solid and dashed curves 
represent model isochrones for pre-main-sequence stars
from \citet{1999ApJ...525..772P} corresponding to 1 Myr and 0.3 Myr, respectively. The 
slanted dotted lines are the reddening vectors for 3 and 0.1 M$_\odot$ 
pre-main-sequence stars corresponding to 1 Myr isochrone. The crosses denote the locations 
of 0.2 and 0.4 M$_\odot$ on the 1 Myr isochrone. Sources that lie above 
the reddening vector for 3 M$_\odot$ are likely luminous massive star candidates. 
In the colour-colour 
diagram, the locii of the main-sequence stars and giants are shown by the 
solid lines (thin and thick), respectively. The short-dashed and dotted lines 
represent the loci of classical T-Tauri stars and Herbig Ae/Be stars, respectively. 
The three parallel dash-dotted lines follow the reddening vectors of giants, main-sequence stars (or dwarfs) and T-Tauri stars. The long dashed line shown is drawn tangentially 
to the turn-off point of the main-sequence locus and used to estimate the reddening 
of the sources (see text for more details).
}
\label{cccm}
\end{figure*}

In order to study the stellar populations towards this region, we have 
constructed the colour-magnitude (CM: $J-H$ vs. $J$) and colour-colour (CC: $H-K$ vs. 
$J-H$) diagrams. For this, we have considered point sources within a circular region of radius $1.3\arcmin$ 
($\sim0.9$~pc) centered on the IRAS peak 
($\alpha_{2000}$ = $17^h$ $29^m$ $16.7^s$, $\delta_{2000}$ = -36$^{\circ}$ 40$\arcmin$ 13$\arcsec$). 
The circle enclosing this region is shown in Fig.~\ref{jhk} (larger circle).
This radius has been selected based on the fact that this
circular region encompasses the radio, millimetre and most 
of the MIR and NIR diffuse emission. We have found a total of 637 
$K_s$ band point sources within this circular region, 261 of which are detected 
in all the three $J$, $H$ and $K_s$ bands and having errors less than 0.1 mag. The CM and CC 
diagrams of this sample of 261 sources are shown in Fig.~\ref{cccm}. 
In the CC diagram, the locii of the main-sequence stars, giants, classical
T-Tauri stars \citep{1997AJ....114..288M} and Herbig Ae/Be stars 
\citep{1992ApJ...393..278L} are shown. The reddening vectors of the main-sequence stars, giants, T-Tauri stars 
are also depicted. All the magnitudes and curves are in the BB system.

We find 61 infrared excess sources which are identified as those lying to the right of 
the reddening vector drawn from the bottom of the main-sequence curve as well 
as those lying in the T-Tauri and HeAeBe zones in the CC diagram. These are 
shown as open circles in the figure. The asterisks represent three young early B or late O type stars whose 
spectral types have been determined from NIR spectroscopy by \citet{2005A&A...440..121B, 2006A&A...455..561B}. They are labelled as 
IR1 (O9-B1/B1-B2), IR2 (early B) and IR3 (O9-B2) for convenience. The details of their positions, magnitudes and spectral types 
are given in Table~\ref{IR}. They represent the brightest objects in the cluster. Of these three, only IR1 and IR2 show infrared excess. 
The rest of the sources in the sample (having no infrared excess or whose spectral types are not known) are represented as dots. 

A total of 22 WISE sources have been detected in our circular region of interest, of which 16 have NIR counterparts. 
We have searched for WISE counterparts to NIR excess sources as well as IR1, IR2 and IR3. Only three NIR excess 
sources have WISE counterparts, of which one is detected in only $W2$ band of WISE. This is possibly a spurious association 
since the NIR source is detected in all the three $JHK_s$ bands. We are therefore left with two 
NIR sources having WISE counterparts. We would like to emphasize here that since the angular resolutions 
of IRSF and WISE are considerably different ($1\arcsec$ and $9\arcsec$, respectively), the association is by no means 
complete. In other words, there are likely to be other NIR excess sources having WISE counterparts but 
we are unable to isolate them due to the angular resolution and sensitivity of WISE images.

In order to get an 
estimate of the extinction suffered by these NIR sources, we have 
dereddened all the sources in the CC diagram along the reddening curve up to a line
drawn tangentially to the turn-off point of the main-sequence locus, shown by the long-dashed line
in Fig.~\ref{cccm}~(right). A histogram of these extinctions shows two marked peaks: the first set ranging from $A_V\sim0-8$ mag and 
peaking at $A_v=0$ mag and the second set ranging from $A_V\sim9-20$ mag (peak $\sim10$ mag) with few sources having extinction beyond 30 magnitudes. 
These two distributions are very likely to represent the foreground sources and the cluster, respectively.

\subsection{$K_s$ band luminosity function}

\begin{figure*}
\hskip -1.cm
\includegraphics[scale=0.45]{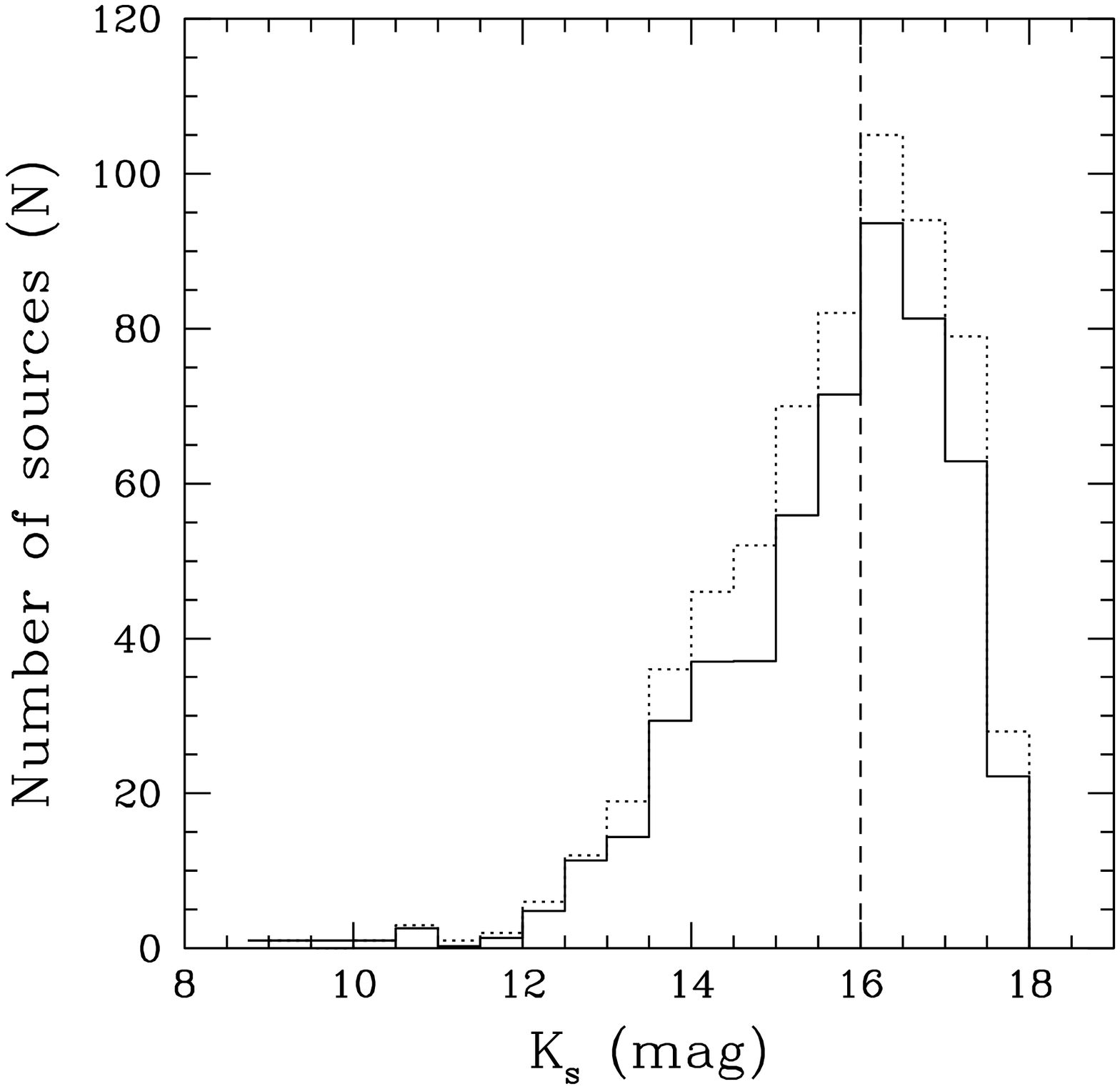}
\includegraphics[scale=0.45]{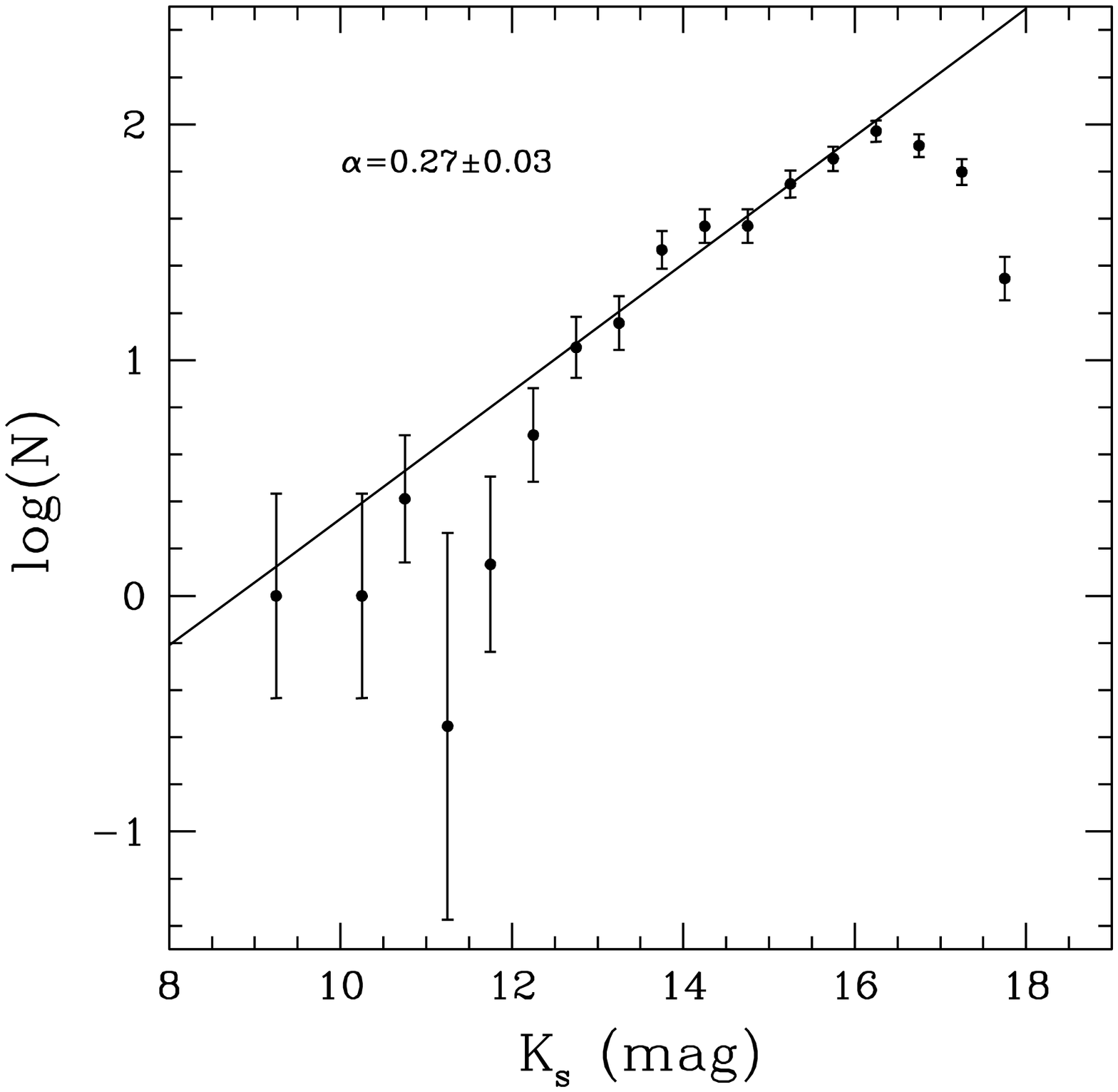}
\vspace*{-3mm}
\caption[KLF of G351]{The corrected K$_s$-band luminosity function (KLF) for the embedded 
cluster associated with G351.63-1.25 is shown as a solid line histogram (left). The 
dotted line histogram is the luminosity function without correction for contamination (foreground 
and background sources) and the dashed vertical line represents the 90\% completeness limit. The KLF is also shown as 
logarithm of number of stars (N) versus the K$_s$ magnitude (right). The filled circles are the observed number of stars 
in a given magnitude bin while the error on this number is given by $\sqrt{N}$. The 
straight line is the least-square fit to the data points in the K$_s$ magnitude range 11.5 -- 16.0.
}
\label{klf}
\end{figure*}

The $K_s$ band luminosity function (KLF) has been constructed and analysed for the embedded cluster 
associated with G351.63-1.25. The KLF can be written as a power law of the form 
$$\frac{dN(K_s)}{dK_s}\propto 10^{\alpha K_s} $$ 
where $\frac{dN(K_s)}{dK_s}$ represents the number of stars per unit magnitude bin in $K_s$ 
and $\alpha$ is the slope of the power law.
In order to obtain the KLF, it is important to get an estimate of the foreground and 
background sources (i.e. sources not associated with the cluster). An estimate of this 
contamination has been obtained using simulated model counts as well as counts from a 
nearby control field. 
For the simulated model counts, we used the Besan\c{c}on model of population synthesis \citep{2003A&A...409..523R}. The star counts
 in the direction $l=351.6^{\circ}$, $b=-1.3^{\circ}$ were 
synthesized using stars in the thin disk, thick disk, stellar halo and outer bulge. The 
fraction of foreground sources in various $K_s$ magnitude bins have been obtained using 
$A_V=0$~mag for distances $d<2.4$~kpc. For the background contribution, we have used model 
counts assuming $A_V=20$~mag and $d>2.4$~kpc. $A_V=20$~mag has been considered based on the 
observations that the extinction in the cluster region is $A_V\sim 10-20$ mag. 
The fraction of foreground and background sources with respect to the total number of stars 
(determined from the model) was multiplied with star counts in the control field to 
get a more accurate estimation of the contamination in the cluster region. This is because 
the number of background sources as estimated from the control field would not have suffered 
extinction due the nebulous region in this star forming region, G351.63-1.25. This has, therefore, been corrected by using the simulations. 
The control field has been selected to be a circular region to the west of the nebula, and centered on 
 $\alpha_{2000}$ = $17^h$ $29^m$ $04.15^s$, 
$\delta_{2000}$ = -36$^{\circ}$ 40$\arcmin$ 30.3$\arcsec$. The background and 
foreground contamination as estimated from the control field is subsequently subtracted 
from the cluster region in order to obtain the KLF. The KLF of the embedded cluster 
is shown in Fig.~\ref{klf}~(left) along with the luminosity function without 
background and foreground correction. We have also estimated the completeness limit of the $K_s$ band by adding artificial stars 
in the observed image. The observations are complete (90\%) to the level of 16.0 mag in this band.

The logarithm of the number of stars as a function of the $K_s$ magnitude for the embedded cluster in this region 
is shown in Fig.~\ref{klf}~(right). We have fitted a linear function to the $\log(N)$ 
versus $K_s$ distribution using least-squares method. The 
best-fit power law is shown as a solid line in Fig.~\ref{klf}~(right). We find $\alpha=0.27\pm0.03$ for the 
cluster using the $K_s$ magnitude bins 11.5 - 16 mag with a bin-size of 0.5 mag. The fitting has been 
carried out after taking into consideration, the statistical error on the number of stars in each bin. The value
 of the power law slope is lower 
than the typical values (0.32 -- 0.38) for other young clusters \citep{1993ApJ...408..471L,1993ApJ...407..657C}. 
However, it has been found that for very young clusters with pre-main-sequence objects, the KLF slope is flatter. 
For instance, \citet{2004ApJ...616.1042O} have obtained $\alpha$ to be $\sim 0.28$
for NGC~7538, whose age is estimated to be $\sim 1$ Myr. Some examples of embedded clusters where even 
lower values of slope have been reported are: W3-IRS5 
\citep[0.18 and 0.17 by][respectively]{1996A&A...307..775M, 2004ApJ...608..797O} with an 
estimated age of $\sim0.3$ Myr. Another example is the case of $\rho$ Ophiucus 
dark cloud \citep[0.17;][]{1992ApJ...395..516G} with ages ranging from $0.1-1$~Myr.

\subsection{Mass spectrum and Age estimate}

The slope of the KLF is expected to increase with age \citep{1992ApJ...395..516G}. The KLF slope can, therefore, 
be used as an age indicator. If we assume similar mass-to-luminosity relation as the young cluster NGC~7538 
\citep[KLF slope $\alpha=0.28$,][]{2004ApJ...616.1042O}, 
then the age of the young embedded cluster in G351.63-1.25 region ($\alpha =0.27$) is consistent with $\sim$ 1 Myr. 
The slope of the initial
mass function (IMF), $\gamma$, is related to the slope of KLF, $\alpha$, through the slope of mass-to-luminosity 
relation $\beta$ as $\alpha=\gamma/(2.5\beta)$ \citep{1996A&A...307..775M}, and hence the IMF slope
depends on the value of $\beta$ used. For $\beta=1$ \citep{1992ApJ...384..212S,1993ApJ...408..471L}, we get 
$\gamma = 0.68$. However, $\beta=1$ is used mostly for stars 
at the lower mass end of the IMF, i.e. G - M stars for a 1 Myr cluster \citep{1996A&A...307..775M}.  If we 
use $\beta = 2$, generally used for a larger and higher mass range (O - F stars) at 1 Myr
\citep{2004AJ....128.2942B}, then we get $\gamma = 1.35$. This matches the standard Salpeter IMF which is given by 
$\gamma = 1.35$.  
A shallower value of $\beta$ gives a flatter slope. A literature search shows that for a 1 Myr cluster associated with
 NGC~7538, \citet{2004AJ....128.2942B} obtain a steeper value of $\gamma\sim1.58$ using $\beta = 2$. 
For younger and older populations within the same cluster, i.e. NGC~7538, \citet{2004ApJ...616.1042O} obtained 
$\alpha = 0.27 - 0.33$. This yields $\gamma = 1.35 - 1.65$, if 
we consider $\beta=2$. Another example is that of 1 Myr Trumpler clusters in Carina Nebula, where 
$\gamma = 1.30 - 1.40$ has been obtained \citep{2007ApJ...667..963S}. 
For younger sub-clusters (0.2-0.3 Myr) associated with Sh~2-233IR, flatter slopes of IMF ($\gamma\sim0.5$) have 
been obtained, \citep{2010ApJ...720....1Y}. 
Given that the slope of the KLF of G351.63-1.25 is similar to clusters of age 1 Myr
 for a given mass-to-luminosity relation, 
we can say that the embedded cluster associated with G351.63-1.25 is very young in nature, whose age 
is compatible with 1 Myr. It is important to note that this method of age estimation using the luminosity function 
is an indirect one. The $J-H$ vs. $J$ CM diagram of the cluster field along with the isochrones corresponding to the ages
 0.3 Myr and 1 Myr is shown in Fig.~\ref{cccm}~(left). The isochrones 
correspond to those of \citet{1999ApJ...525..772P}. 
From Fig.~\ref{cccm}~(left), we see that if we consider the 1 Myr isochrone, 
then the young stellar objects have masses higher than $\sim0.1$ M$_\odot$. In other words, 
our dataset is able to probe young stellar objects of masses up to $\sim0.1$ M$_\odot$. 

Another age indicator is the fraction of NIR excess stars in a cluster. This is because the disk and/or 
envelope associated with a pre-main-sequence star starts becoming optically thin with age. This NIR excess fraction method has been used to estimate 
the ages of young embedded clusters \citep{2003ARA&A..41...57L}. For young clusters whose age is 1 Myr, the fraction of NIR excess stars (based on
JHK) is found to be $\sim50$\% \citep{2000AJ....120.3162L, 2000AJ....120.1396H} to $\sim65$\% \citep{2001ApJ...558L..51M}. 
In the case of Taurus dark clouds of age $1-2$ Myr, the NIR fraction is estimated to be
$\sim40$\% \citep{1995ApJS..101..117K} which decreases to  $\sim20$\%
for older clusters of age $2-3$ Myr \citep{2005ApJ...629..276T}. To estimate the NIR excess fraction, it 
is important to know the contamination in the cluster region due to foreground and background sources. We, again, use the 
simulated counts from the Besan\c{c}on model and the star counts in the control field to estimate the 
fraction of foreground and background contaminating stars, which is 35\%. After correcting
for the photometric completeness as well as the foreground and background
star contamination, the fraction of the NIR excess stars is estimated to be $43$\% indicating an upper age limit of 
$1-2$ Myr. This is compatible with the age estimate obtained earlier by the KLF method.

\section{[Ne\,{\sevensize\bf II}] Emission}

The MIR \neii at 12.82 $\mu$m is a fine-structure line excited by ultraviolet photons of energy $>21.6$ eV and hence detection 
of this line traces far UV photons very close to the young massive star. 
The \neii line emission along with physical parameters derived from the radio emission can be used to estimate
the effective temperature of the ionizing star \citep{1998ApJ...507..263W}.
In the IRAS-LRS spectrum of G351.63-1.25, a strong emission line at 
12.8 $\mu$m corresponding to \neii is observed. The \neii 
line intensity is $F_{[NeII]}\sim 8.2\times10^{-17}$ W cm$^{-2}$. This has been determined
 by fitting a gaussian function to the line and integrating the area under the 
curve after subtraction of a polynomial baseline. 

 Although the \neii line is from a wide 
aperture of $5\arcmin$ 
(corresponding to the IRAS-LRS detectors), we estimate the spectral 
type of the exciting source assuming that this emission arises due to a single 
massive star associated with S2. 
 This is because the formalism considered here uses \neii emission 
due to a single ZAMS star to estimate its effective temperature.
The Ne$^+$ abundance ($Ne^+$/$H^+$) is obtained using the following
equation for the case when the electron density is below 
the critical density ($n_{crit} = 3.6\times10^5$ cm$^{-3}$), \citep{2000ApJ...541..779T}:

$$\frac{Ne^+}{H^+}=\frac{F_{[NeII]} {\rm (W\; cm^{-2})}}{2.3\times10^{-9}\;\Omega_b\;T_e^{-1/2}\,e^{-(hc/\lambda kT_e)}\;(EM)}$$

where $F_{[NeII]}$ is the \neii line flux density, $\Omega_b$ is the solid angle of the \neii emitting region in
Sr, $T_e$ is the electron temperature in K, $\lambda$ in the Boltzmann factor is the wavelength of the \neii line 
(12.81 $\mu$m in this case), and $EM$ is the emission measure in cm$^{-6}$ pc. The electron density obtained 
for S2, $n_e = 1.2\times10^4$~cm$^{-3}$ is an order of magnitude lower than the critical density. Further, we assume that the 
spatial distribution of \neii line emission overlaps the compact core of the 
ionised gas implying $\Omega_b\sim12''$. For this region, we find Ne$^+$ abundance to be 
$\sim4.5\times10^{-5}$. \citet{2000ApJ...541..779T} have calculated (and plotted) the
Ne$^+$ abundance as a function of the effective temperature of 
the ionizing star from the \hii~region model using the {\it CoStar} 
stellar atmosphere model of \citet{2000ApJ...541..779T}. Using this, we find that our
Ne$^+$ abundance
 corresponds to a stellar effective temperature of 
$\sim36000$ K, i.e a star of spectral type O8.5-O8. 
This is later than the spectral type of the ionising star obtained from radio 
measurements (O7.5). This could be explained on the 
basis of the size of the Ne$^+$ emitting region. We have assumed that 
the size of the \neii emitting region coincides with the compact radio peak. 
However, it has been found that for the relatively massive stars causing 
ionisation, the size of the Ne$^+$ 
abundance shows a dip at the location of the radio peak, due to larger 
Ne$^{++}$ abundance \citep{2000ApJ...541..779T}. 
These authors also find that the spectral type of the ionising star obtained from the Ne$^+$ abundance measurements is 1-2 subclasses later 
than the spectral type obtained from radio measurements. 

\section{The multiwavelength scenario}

In this section, we analyse and interpret our results in different wavebands in order to understand various aspects of the star 
formation process ocurring in this region.

\subsection{Nature of S2}

Among the six ionised clumps listed in Table~\ref{radsource}, the brightest is S2 
 whose excitation is consistent with a ZAMS star of radio spectral type O7.5. The radio morphology of S2 at 1280 MHz is 
not point-like. This is confirmed from the high angular resolution ($1.5\arcsec$) map at 8.64 GHz by \citet{1998MNRAS.301..640W}. 
In their map, S2 displays an irregular morphology comprising near equal intensity five peaks, with a typical separation of $1-3\arcsec$. Assuming 
that the emission is optically thin and each radio peak represents a ZAMS O/B star, this would mean that a cluster of OB stars exist within a region 
of size 0.07 pc. On the other hand, these could be high density clumps ionised externally. In that case, since they are of nearly equal brightnesses, 
optical depth must be playing an important role. While the 8.64 GHz map images only the finer details, the 1280 MHz image shows more nebulous
components with shoulders in emission ($19\arcsec\times23\arcsec \sim 0.2\times0.3$~pc$^2$). We can therefore consider S2 as an ultracompact 
\hii~region 
inside a compact component which in turn is embedded in a more extended emission component \citep{2001ApJ...549..979K}.

S2 does not coincide with any known infrared point source within a search radius of $3\arcsec$ ($\sim 0.03$~pc), 
as investigated using IRSF and MSX. There is 
an offset between the radio and MIR peaks. This implies that ionising source(s) is deeply embedded and suffer extinction even in 
the MIR. This is clearly evident from the WISE images in Fig.~\ref{wise}. This hypothesis (high extinction towards S2) is further 
validated by the presence of a dense molecular (HC$_3$N) core (with virial mass upper limit is 940 M$_\odot$) by \citet{2004AJ....128.2374S}.  
The source closest to S2 ($3.2\arcsec$) in the NIR is a very `red' source detected in H and K$_s$ bands only ($H - K = 3.52$ mag; 
$K_s$ = 13.08 mag) which has also been detected at 4.5 $\mu$m. This source is adjacent to the west of a high density filamentary 
structure running vertically, which is visible in the colour-composite $JHK_s$ image in Fig.~\ref{jhk}. The closest  
star detected in all three $JHK_s$ bands in the vicinity of S2 is located $\sim6.8\arcsec$ away and has NIR extinction larger than $A_V\sim20$ mag. 
The massive young stellar object, IR2, is at a distance $8.6\arcsec$ from the radio peak. \citet{2006A&A...455..561B} carried out VLT K-band 
spectroscopy of IR2 (referred as 17258nr593 in their paper) and classified it as an early B star (having 16 magnitudes of visual 
extinction). IR2 has an infrared excess as seen in the CC diagram. The bright MIR young stellar 
object detected at 3.4~$\mu$m WISE and 8.3~$\mu$m MSX images lies at the western edge of S2 and is probably, responsible for the small shoulder 
in radio emission seen towards the west of S2 (Fig.~\ref{wise}~left). 
The presence of massive young stellar objects like IR2, the infrared excess sources as well as the sources detected only 
in H and K$_s$ bands lends credibility to our hypothesis that S2 is excited by a cluster of embedded young massive objects. 
The compact ionised component of S2 is therefore being powered by UV photons escaping from the UCHII region as well as by 
other massive embedded young stellar objects (like IR2). 
This concurs with the conclusion of \citet{1994ApJS...91..659K} that many ultracompact \hii~regions are powered by a cluster of stars.

\subsection{Do the other ionised clumps harbour stars?}

We first search for NIR counterparts to radio peaks (other than S2) within a search
radius of $3\arcsec$ considering the resolution of radio map at 1280 MHz. We do not find any NIR counterpart to S1, S3, S4 and S6.
For S5, the closest NIR source is $0.9\arcsec$ away. 
However, this is neither an infrared excess source nor an early spectral type 
object (based on the CM diagram). It is very likely to be a foreground source 
along the same line-of-sight. The young stellar object IR3 is located $\sim10\arcsec$ away from S4.
\citet{2004PhD...A.Bik} has also carried out K band spectroscopy of IR3 (17258nr378 in his work)
and infer its spectral type as O9-B2. It appears unlikely to be a massive young star as we do not observe compact radio emission coincident with IR3. 
Neither do \citet{1998MNRAS.301..640W} find any compact source here. Therefore, the radio emission indicates that the spectral type of IR3 is 
consistent with B2-B1. The NIR source closest to S4 is $3.7\arcsec$ away and detected only in $K_s$ band ($K_s$ = 15.25 mag). 
With the given data, it is difficult to ascertain which young stellar object is responsible for the radio peaks, 
but it is likely that the more massive ones among these are contributing to the radio emission. 

We now explore the hypothesis that these high density ionised regions are externally ionised clumps located in an elongated \hii~region  
created by a group of young massive stars located in S2. Note that the peak flux densities are lower than that of S2 by a factor of 5 or more. 
The flux densities  
 at the location of these clumps based on solid angle subtended by them from the ionising star(s) are estimated. For clumps other than S5, the
 derived values are much lower than observed flux densities (7-30\%). For S5, this ratio is higher $\sim80$\%. This would imply that most of 
these ionised clumps have an embedded ZAMS exciting star. However, this result must be treated with caution as there are two opposing factors 
that have been ignored here. First, these estimates were obtained assuming an absence of intervening medium, in which case the ratios would go 
lower. On the other hand, we have assumed optically thin emission, for S2 as well as the ionised clumps. However, we have seen that S2 is 
optically thick and high optical depth for the other clumps cannot be ruled out. Consequently, the flux densities cannot be compared directly 
and a map with optically thin emission from all the clumps would enable  a better comparison. In addition, we do not find NIR 
counterparts to these ionised clumps; and the morphology of the ionised clumps are relatively less compact, with the exception 
of S3 and S5. At higher radio frequencies, the emission from these clumps are resolved out at by an extended array configuration 
\citep{1998MNRAS.301..640W}. These could, therefore, be more evolved {individual \hii~regions in case they harbour stars}. 
However, since we do not observe the associated exciting stars in infrared, we are inclined to take the view that these are 
high density clumps in the extended emission of the ultracompact \hii~region.  

\subsection{Star formation activity}

The radio continuum emission at 1280 and 610 MHz traces extended as 
well as compact emission in the region associated with G351.63-1.25. 
The extended emission from ionised gas is elongated along the northeast-southwest direction. On both sides of the elongated emission, we 
observe steep gradients in brightness distribution. This signifies that this region is
ionisation bounded on both sides suggesting an encounter with the ambient molecular gas during the expansion phase of the \hii~region.
If we compare the morphology of the diffuse emission in the radio and 
MIR, we perceive that the radio contours follow the 
MIR diffuse emission from warm dust. This is clearly visible in 
Fig.~\ref{wise}~(left) towards the north of S2 
as well as along the elongation. Although not shown here, the high resolution 8.0 \micron {\it Spitzer}-IRAC image (which 
covers only the northern part of this diffuse emission) resolves the diffuse 
emission into filamentary structures. High extinction filamentary structures are also seen in the NIR images 
entwined with the nebulous emission. This indicates that this region is highly inhomogeneous implying a variation in extinction.

\begin{figure*}
\hskip -1cm
\includegraphics[scale=0.45,angle=-90]{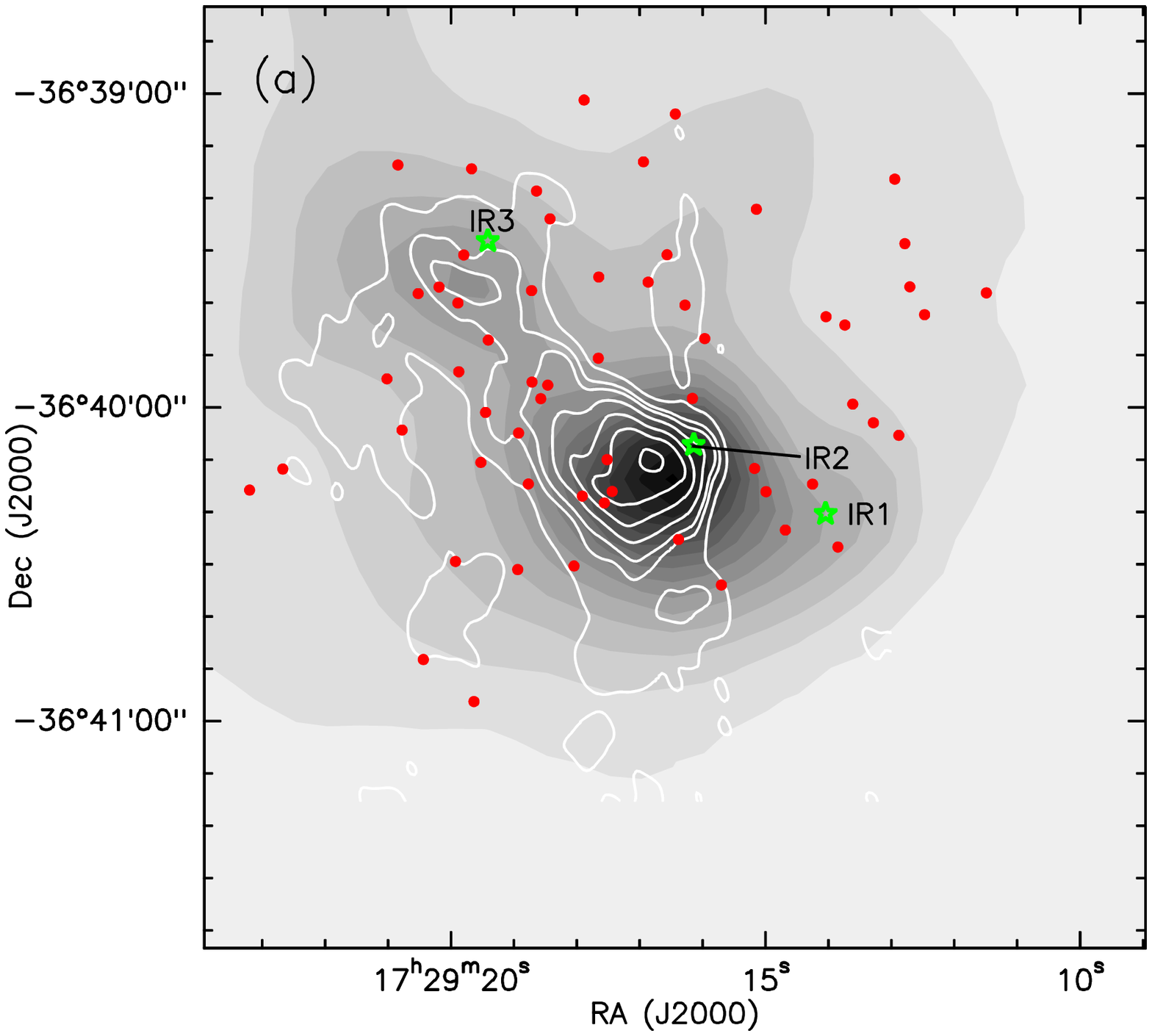}
\includegraphics[scale=0.45,angle=-90]{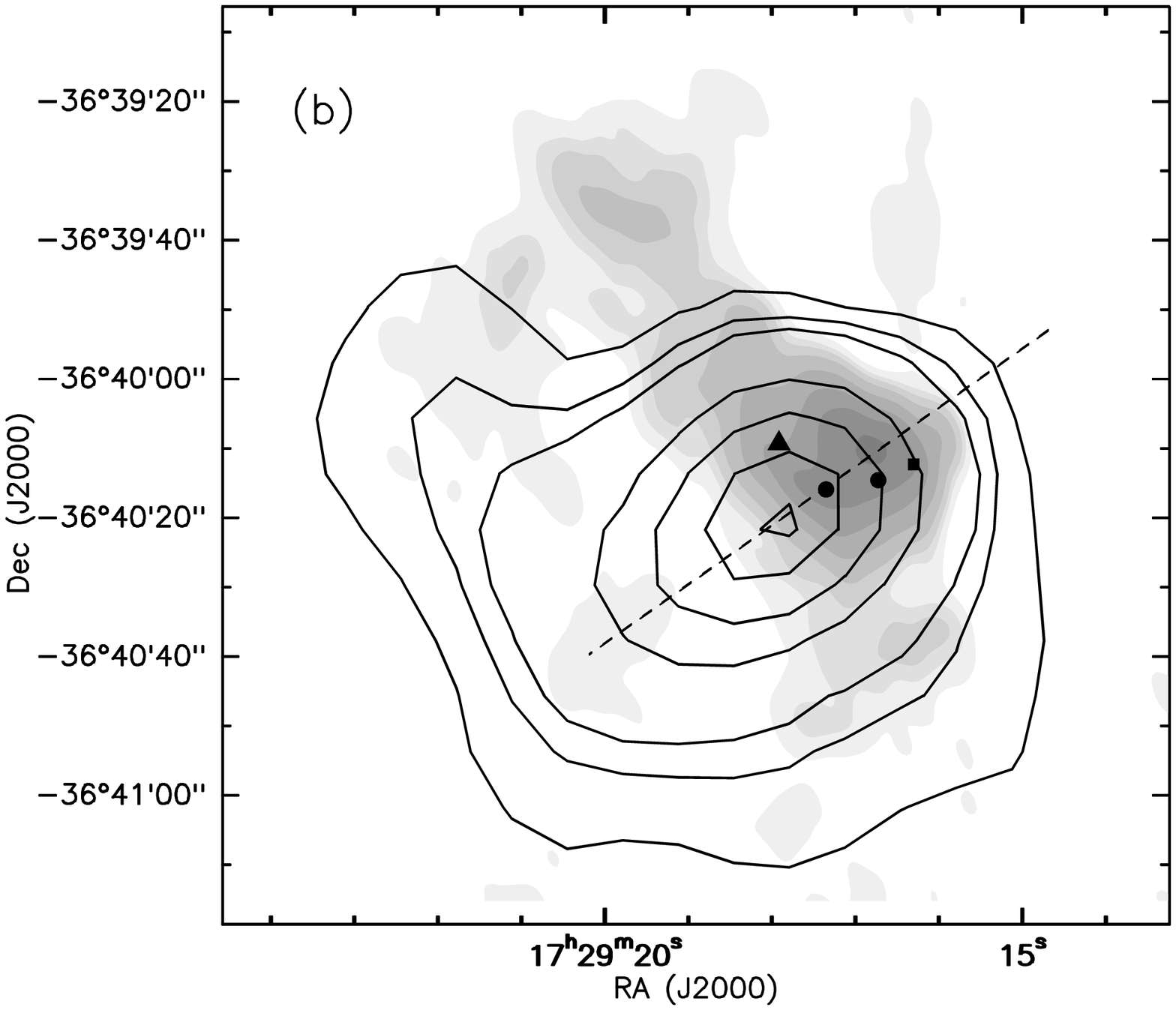}
\caption[All]{ (a) Grayscale emission from warm dust (MSX 8 \micron) in the vicinity of G351.63-1.25 with levels ranging from 
$2.5\times10^{-7}$~Wm$^{-2}$Sr$^{-1}$ to $5\times10^{-4}$~Wm$^{-2}$Sr$^{-1}$ 
in steps of  $1.8\times10^{-5}$~Wm$^{-2}$Sr$^{-1}$. The solid line contours 
represent the 1280 MHz radio continuum emission, the contour levels are same as 
those shown in Fig.~\ref{wise} (left). The asterisks represent young stellar objects whose near-infrared spectral type has been inferred based on 
other spectroscopic studies and the circles represent infrared excess sources. (b) Grayscale is 1280 MHz radio emission, while the contours 
represent 1.2 mm emission with contour levels ranging from 
0.5 to 14 Jy/beam in steps of 1.5 Jy/beam; beam is $24\arcsec$. The triangle represents the position of Class I methanol maser. The filled square and 
circles represent emission peaks at 3.4 and 4.6 $\mu$m from WISE, respectively. The 
dashed line represents the direction of thin flat molecular cloud (see text for more details).}
\label{all}
\end{figure*}

The emission from cold dust at 1.2 mm from \citet{2004A&A...426...97F} is 
shown in Fig.~\ref{all}~(right) as contours. The millimetre emission shows 
a core with peak flux of 13.8 Jy/beam, where the beam size is $24\arcsec$. The total mass of this cold core 
estimated from millimetre emission is 1400 M$_{\odot}$. 
A core dust temperature of 42 K is obtained based on a grey body fit to the fluxes 
at the far-infrared wavelengths by \citet{2004A&A...426...97F}.  The dust optical depth at 1.2 mm 
is estimated to be $3.7\times10^{-3}$ which translates to an optical 
depth of 0.3 at 100 \micron if we consider the extinction from silicate 
grains of \citet{1983A&A...128..212M}. 
This region has been detected as a compact core in the 2 and 3 mm QUaD survey \citep{2011ApJS..195....8C}
with flux densities $37.6\pm0.3$ (beam $\sim3.8\arcmin\times1.8\arcmin$) and $30.9\pm0.2$ Jy 
(beam $\sim4.0\arcmin\times2.3\arcmin$), respectively. While the 3 mm emission has non-negligible contribution 
from the free-free emission to the dust emission, we assume this is not the case for the 2 mm emission. It is 
difficult to estimate the mass of such a cold clump, using the 2 mm emission, as the values of dust temperature and
 opacity are not known 
accurately. If we extrapolate the formalism of \citet{1983QJRAS..24..267H} to millimetre wavelengths, we get the dust 
opacity as $\kappa_{2\,mm} = 0.01$~g/cm$^2$. This leads to a dust mass of 360~\msun for a dust temperature of 30K. 
This is an order of magnitude larger than the mass of dust emitting at 1.2 mm \citep{2004A&A...426...97F}. 
Considering that the resolution is poorer at 2 mm, it is likely that other cold clumps in the vicinity could be 
contributing to this emission. A comparison of the morphology of the warm and cold dust distributions 
from MIR and 1.2 mm maps respectively, shows that the 
emission at different wavebands peak at different locations. The millimetre emission 
peaks at ($\alpha_{2000}$, $\delta_{2000}$) = ($17^h$ $29^m$ $18.11^s$, -36$^{\circ}$ 40$\arcmin$ 21.0$\arcsec$) towards the 
south-east ($\sim20\arcsec$) of the peak emission from warm dust and ionised gas (separation 
$\sim19\arcsec$). This is evident from Fig.~\ref{all}~(right). 

A search for masers in this region shows that a methanol maser has been 
detected here \citep{1994MNRAS.268..464S,2000MNRAS.317..315V}. The triangle in Fig.~\ref{all}~(right)
shows the location of the peak methanol maser emission. This is a Class I methanol maser detected at 44 and 95 GHz.
\citet{2004ApJS..155..149K} find a large number of these masers in relatively close association with other 
massive star formation signposts like water masers and \hii~regions. In their 
VLA survey of star-forming regions, they find that these masers are typically 
offset by $\sim0.2$ pc (median value) from the star formation signposts. In 
the case of G351.63-1.25, the maser is offset from the peak radio emission by 0.2 pc ($\sim14\arcsec$). It is believed that 
Class I masers are associated with shock fronts, indicating the interface 
of interaction between mass outflows and dense ambient material \citep{1990ApJ...364..555P}. Note that the observations of 
\citet{2004AJ....128.2374S} point towards
tentative detection of CO and SiO outflows from this region. The shock 
front exciting the maser is probably due to the interface (along the elongation) of the ionised 
gas/outflow and the cold millimetre clump. The absence of radio 
and infrared emission peaks near the millimetre emission peak indicates that formation of stars is either in a very early stage 
or has not yet begun.

The investigation of G351.63-1.25 at different wavelength bands ranging from 
NIR to millimetre wavelengths, leads to the following scenario. 
Based on evidences relating to (a) morphology of S2, (b) absence of infrared source corresponding to S2 peak position, 
(c) presence of young stellar objects within nebulosity of S2 (d) detection of a dense molecular gas (HC$_3$N) core close to S2 
(e) presence of Class I methanol maser, and (f) tentative detection of CO and SiO outflows, we can say that the G351.63-1.25 comprises a 
group of embedded massive young stellar objects responsible for ionisation of the compact source 
S2. Regarding other ionised clumps, we are inclined to take the view that these are externally ionised by the central cluster, based on 
(a) absence of infrared counterparts, and (b) optical depth effects. Further high frequency radio observations, 
sampling emission from the ionised clumps, are needed in order to ascertain that they host massive stars. 
  
The morphology of the \hii~region is elongated with ionisation bounds on either side of the elongation and density bounds along the 
elongation direction, similar to a bipolar \hii~region \citep{2001ApJ...549..979K, 1984A&A...135..261F}. Note that the elongation is 
perpendicular to the direction of the line joining S2 and the millimetre peak, shown by dashed line in Fig.~\ref{all}~(right). Further, 
the millimetre cloud is elongated in this direction (dashed line). This suggests that the molecular cloud is dense along this 
direction with massive stars being formed here, vis-a-vis S2. The peak brightnesses at 3.4 and 4.6~$\mu$m due to warm dust
also lie in this direction lending support to this hypothesis. We propose a scenario where the star formation 
proceeds in the flat molecular cloud while the expansion of the \hii~region occurs away from it, in a direction 
perpendicular to the molecular cloud. Multiwavelength observations indicate the presence of massive stars, cold dust emitting millimetre emission 
as well as hot dust emitting near- and mid-infrared emission in this flat molecular cloud. As these lie at different 
locations within the flat molecular cloud, it implies that G351.63-1.25 is a region that displays different 
evolutionary stages of star formation. 

The bipolar radio morphology in extended emission can be explained in terms of the champagne flow model 
\citep{1978A&A....70..769I, 1979A&A....71...59T}. The champagne flow model explains the shape of a \hii~region to be
 due to the density gradient of the molecular material in the vicinity of the newly formed massive star. 
The present scenario of flat molecular cloud has been modelled by \citet{1979ApJ...233...85B} where the ionisation
 front breaks out through two opposing faces of the same cloud. A star located within a disk-like flat molecular cloud 
produces an \hii~region which breaks through the cloud in both directions in a double cone-like structure, the opening
 angle of which widens over time. In the 
case of G351.63-1.25, if we assume that the exciting cluster of stars is located in a thin, flattened 
molecular cloud (indicated by the millimetre cloud, infrared peaks and maser spot), then the expansion of the \hii~region would lead to a 
bipolar type of morphology in a direction perpendicular to the flat cloud as the density gradient is maximum here. The high 
density ridge (towards the lower part of the fan-shaped nebulous region) is rougly in the same direction as the flat molecular cloud 
(dashed line in Fig.~\ref{all}~(right)). If this is part of the same thin, flat molecular cloud, then the geometry implies that the thin cloud 
is not fully edge-on (i.e. inclination angle $<90^\circ$) to our line-of-sight. This would also explain the asymmetry in 
length of the lobes of the \hii~region. However, high angular 
resolution molecular line imaging observations of the ambient molecular cloud will be required to confirm this.

\section{Conclusions}

Based on the multiwavelength (radio, infrared and millimetre) investigation of the 
star-forming region associated with G351.63-1.25 presented here, we come to the following conclusions. 

\begin{enumerate}

\item The radio map of the \hii~region at 1280 MHz comprises of six high density ionised clumps embedded in diffuse emission. The brightest clump at S2 
is an ultracompact \hii~region with the electron temperature $\sim 7647\pm153$ K and emission measure 
$\sim 2.0\pm0.8\times10^7$~cm$^{-6}$pc. The equivalent ZAMS spectral type is estimated to be O7.5.

\item The NIR broad band images in J, H and K$_s$ reveal the presence of 
fan-shaped nebulous emission as well as high extinction filamentary structures. The stellar component is
probed using
colour-magnitude and colour-colour diagrams. These have been used to find the infrared excess sources, 
associated with the embedded cluster. The log-normal slope of the KLF of the embedded cluster after removing the 
contamination due to foreground and background sources is $\sim 0.27\pm0.03$, indicating the youth of the cluster. 
The fraction of the NIR excess stars is estimated to be 43\% indicating an upper age limit of 1 - 2 Myr.
Based on KLF as well as NIR excess fraction, we believe that age of the cluster is compatible with $\sim1$ Myr. 

\item The MIR images from WISE and MSX show diffuse emission that matches the ionised gas distribution very well.

\item The ultracompact component S2, does not have an infrared counterpart within 
$3\arcsec$ of the radio peak at 1280 MHz. Further, based on the morphology of S2 (not point-like), the presence of 
other young stellar objects within the nebulosity of S2, the detection of dense molecular gas core close to it, 
the presence of Class I methanol maser  as well as tentative evidence of SiO outflows, it is very likely that S2 is 
ionised by a group of massive embedded sources rather than a single source.

\item The ionised clumps seen in the 1280 MHz map are likely to be externally ionised by the central cluster associated with S2 due to absence of 
NIR counterparts, less compact morphology, and optical depth effects at 1280 and 610 MHz.  

\item The warm and cold dust distributions (from MIR and millimetre emission) peak at different locations. The ionised 
emission is elongated indicating ionisation bounds on either side. The elongation is perpendicular to the direction of the line joining the peak 
brightness of ionised gas and cold dust. This is explained using the champagne flow model where star formation 
occurs in a thin, flat molecular cloud and the expansion of the \hii~region happens away from the cloud leading to a bipolar-type morphology 
of extended emission.  

\end{enumerate}

\section*{Acknowledgements}

We thank the staff of the GMRT, who have made the radio observations possible.
GMRT is run by the National Centre for Radio Astrophysics of the Tata Institute
 of Fundamental Research. We also thank the staff of IRSF at S. Africa in joint partnership between the S.A.A.O and Nagoya University of Japan for 
their assistance and support during observations. We thank L. Bronfman for providing 
us the 1.2 mm map of this region. We thank Anne Robin for letting us use her model of stellar population synthesis.

We thank IPAC, Caltech, for providing us the HIRES-processed IRAS products.
This research made use of data products from the Midcourse
Space Experiment. Processing of the data was funded by the Ballistic
Missile Defense Organization with additional support from NASA Office of Space
Science. This research has also made use of the NASA/ IPAC Infrared Science
Archive, which is operated by the Jet Propulsion Laboratory, Caltech, under
contract with the NASA. This publication makes use of data products from the Wide-field Infrared Survey Explorer, 
which is a joint project of the University of California, Los Angeles, and the Jet 
Propulsion Laboratory/California Institute of Technology, funded by the National Aeronautics and Space Administration.

\bibliography{svig}


\begin{table*}
\begin{center}
\caption[Details of radio observations of G351.63-1.25]{Details of the
radio continuum observations carried out for G351.63-1.25 using the GMRT, India.}
\label{radobs}
\vspace{0.2cm}
\begin{tabular}{|l l l l |} \hline \hline
Frequency band (MHz) & 1280 & 610 & 325 \\
\hline
Date of Observation & 29 Oct 2001 & 22 June 2000 & 04 Oct 2003 \\
Primary beam & 26$'.2$ & 54$'$ & 1$^{\circ}.8$ \\
Synthesized beam & $7.0''\times 3.6''$ & $11.7'' \times 8.0''$ & $36.8'' \times
16.2''$ \\
Map noise (mJy/beam) & 2 & 2 & 3 \\
Continuum bandwidth & 16 MHz & 16 MHz & 16 MHz \\
Flux Calibrator & 3C286, 3C48 & 3C48 & 3C286, 3C48 \\
Phase Calibrator & 1626-298 & 1822-096 & 1830-360 \\
\hline
\end{tabular}
\end{center}
\end{table*}



\begin{table*}
\begin{center}
 \caption[Extracted sources from 1280 MHz map of G351.63-1.25]{List of
extracted sources along with their peak and integrated
flux densities obtained from the radio maps of G351.63-1.25 at 1280 and 610 MHz. }
\label{radsource}
\vspace{0.3cm}
\begin{tabular}{|c c c r r  r r|} \hline \hline
Source & RA & Dec & \multicolumn{2}{c|}{1280 MHz} & \multicolumn{2}{c|}{610 MHz}  \\
No.       & (J2000)   &  (J2000)   & \multicolumn{2}{c|}{---------------------------------------} & \multicolumn{2}{c|} {---------------------------------------}
 \\
  &  & & \multicolumn{1}{c|}{Peak Flux} & \multicolumn{1}{c|}{Int flux} & \multicolumn{1}{c|}{Peak Flux} & \multicolumn{1}{c|}{Int flux} \\
 & (h m s) & ($^{\circ}$ $'$ $''$) & (mJy/bm) & \multicolumn{1}{c|}{(mJy)} & (mJy/bm) & \multicolumn{1}{c|}{(mJy)} \\ \hline
S1 & 17 29 16.42 & -36 40 38.0 & $36.0\pm4.5$ & $372.7\pm51.4$ & $56.6\pm2.6$ & $245.8\pm13.7$   \\
S2 & 17 29 16.77 & -36 40 10.4 & $326.5\pm3.4$ & $5894.3\pm65.1$ & $238.6\pm2.6$ & $2345.0\pm27.6$  \\
S3 & 17 29 17.52 & -36 40 49.4 & $23.5\pm4.6$ & $167.8\pm37.6$ & - & - \\
S4 & 17 29 19.61 & -36 39 37.4 & $59.3\pm3.9$ & $846.7\pm59.0$ & $105.9\pm2.6$ & $799.3\pm21.8$     \\
S5 & 17 29 19.91 & -36 39 50.0 & $23.9\pm4.7$ & $148.6\pm33.4$ & - & -  \\
S6 & 17 29 20.96 & -36 39 44.0 & $28.4\pm4.3$ & $359.9\pm59.3$ & $58.5\pm2.6$ & $336.6\pm17.3$  \\ \hline
\end{tabular}
\end{center}
\end{table*}



\begin{table*}
\begin{center}
\caption{Details of  MSX PSC sources in the region associated with 
G351.63-1.25.}
\label{msxpsc}
\vspace*{1cm}
\hspace*{-1.5cm}
\begin{tabular}{|c c c c c c c c|} \hline \hline
MSX PSC & Name$^a$ & $\alpha_{2000}$ & $\delta_{2000}$ & $F_{8}$ & $F_{12}$ &
$F_{14}$ & $F_{21}$\\
designation & & (deg) & (deg) & (Jy) & (Jy) & (Jy) & (Jy) \\ \hline
G351.6423-01.2409 & M1 & 262.3143 & -36.6560 & $2.8\pm0.1$ & $4.1\pm0.2$ & - & - \\
G351.6326-01.2523 & M2 & 262.3195 & -36.6704 & $28.0\pm0.1$ & $97.1\pm4.9$ & $166.1\pm10.1$ & $778.8\pm46.7$ \\
G351.6469-01.2554 & M3 & 262.3325 & -36.6602 & $8.8\pm0.4$ & $23.8\pm1.2$ & $28.7\pm1.7$ & $130.0\pm7.8$ \\ \hline
\end{tabular}
\end{center}

$^a$ Short name used in the present work \\
\end{table*}



\begin{table*}
\begin{center}
\caption{Details of the IRSF-SIRIUS sources in the G351.63-1.25 region with known spectral types based on near-infrared 
spectroscopic studies by \citet{2005A&A...440..121B} and \citet{2006A&A...455..561B}.}
\label{IR}
\vspace*{1cm}
\hspace*{-1.5cm}
\begin{tabular}{|c c c c c c c|} \hline \hline
Source & $\alpha_{(2000)}$ & $\delta_{(2000)}$ & $J$ & $H$ & $K_{s}$ & Spectral\\ 
Name & (deg) & (deg) & (mag) &  (mag) & (mag) & Type \\ \hline
IR1 & 262.30841 & -36.67229 & $12.46\pm0.02$ & $11.24\pm0.03$ & $10.49\pm0.04$ & O9-B1/B2-B3 \\
IR2 & 262.31714 & -36.66863 & $13.91\pm0.03$ & $11.19\pm0.05$ & $9.21\pm0.04$ & early B\\
IR3 & 262.33078 & -36.65780 & $13.29\pm0.03$ & $11.11\pm0.03$ & $10.05\pm0.03$ & O9-B2 \\ \hline
\end{tabular}
\end{center}
\end{table*}


\newpage


\label{lastpage}

\end{document}